\documentclass[twocolumn]{aastex61}
\usepackage{natbib}
\usepackage{graphicx}

\usepackage{soul} 

\received{}
\revised{}
\accepted{}

\shorttitle{$R$-Process Alliance}
\shortauthors{Hansen et al.}

\begin{document}

\title{The $R$-Process Alliance: First Release from the Southern Search for $r$-Process-Enhanced Stars in the Galactic Halo\footnote{This paper includes data gathered with the 2.5 meter du Pont telescope located at Las Campanas Observatory, Chile.}}

\correspondingauthor{Terese T. Hansen}
\email{thansen@carnegiescience.edu}

\author{Terese T.\ Hansen}
\affiliation{The Observatories of the Carnegie Institution for Science, 
813 Santa Barbara Street, Pasadena, CA 91101, USA}

\author{Erika M.\ Holmbeck}
\affiliation{Department of Physics, University of Notre Dame, Notre Dame, IN 46556, USA}
\affiliation{Joint Institute for Nuclear Astrophysics -- Center for the Evolution of the Elements (JINA-CEE), USA}

\author{Timothy C.\ Beers}
\affiliation{Department of Physics, University of Notre Dame, Notre Dame, IN 46556, USA}
\affiliation{Joint Institute for Nuclear Astrophysics -- Center for the Evolution of the Elements (JINA-CEE), USA}

\author{Vinicius M.\ Placco}
\affiliation{Department of Physics, University of Notre Dame, Notre Dame, IN 46556, USA}
\affiliation{Joint Institute for Nuclear Astrophysics -- Center for the Evolution of the Elements (JINA-CEE), USA}

\author{Ian U.\ Roederer}
\affiliation{Department of Astronomy, University of Michigan, 1085 S. University Ave., Ann Arbor, MI 48109, USA}
\affiliation{Joint Institute for Nuclear Astrophysics -- Center for the Evolution of the Elements (JINA-CEE), USA}

\author{Anna Frebel}
\affiliation{Department of Physics and Kavli Institute for Astrophysics and
Space Research, Massachusetts Institute of Technology, Cambridge, MA 02139, USA}
\affiliation{Joint Institute for Nuclear Astrophysics -- Center for the Evolution of the Elements (JINA-CEE), USA}

\author{Charli M.\ Sakari}
\affiliation{Department of Astronomy, University of Washington, Seattle, WA 98195-1580, USA}

\author{Joshua D.\ Simon}
\affiliation{The Observatories of the Carnegie Institution for Science,
813 Santa Barbara Street, Pasadena, CA 91101, USA}

\author{Ian B.\ Thompson}
\affiliation{The Observatories of the Carnegie Institution for Science, 813 Santa Barbara Street, Pasadena, CA 91101, USA}

\begin{abstract}
The recent detection of a binary neutron star merger and the clear evidence for the decay of radioactive material observed in this event have, after sixty years of effort, provided an astrophysical site for the rapid neutron-capture ($r$-) process which is responsible for the production of the heaviest elements in our Universe. However, observations of metal-poor stars with highly-enhanced $r$-process elements have revealed abundance patterns suggesting that multiple sites may be involved. To address this issue, and to advance our understanding of the $r$-process, we have initiated an extensive search for bright ($V < 13.5$), very metal-poor ($\mathrm{[Fe/H]} < -2$) stars in the Milky Way halo exhibiting strongly-enhanced $r$-process signatures. This paper presents the first sample collected in the Southern Hemisphere, using the echelle spectrograph on du Pont 2.5m telescope at Las Campanas Observatory. We have observed and analyzed 107 stars with $-3.13 < \mathrm{[Fe/H]} < -0.79$. Of those, 12 stars are strongly enhanced in heavy $r$-process elements ($r$-II), 42 stars show moderate enhancements of heavy $r$-process material ($r$-I), and 20 stars exhibit low abundances of the heavy $r$-process elements and higher abundances of the light $r$-process elements relative to the heavy ones (limited-$r$).This search is more successful at finding $r$-process-enhanced stars compared to previous searches, primarily due to a refined target selection procedure that focuses on red giants.   
\end{abstract}


\section{Introduction} \label{sec:intro}

For more than sixty years, astronomers have searched for the astrophysical site of the heaviest elements in our Universe -- the elements produced by the rapid neutron-capture process ($r$-process), first described by \cite{burbidge1957} and \citet{cameron1957}. Neutron star mergers (NSMs) have long been proposed as sites for the $r$-process \citep{lattimer1974,rosswog2014,thielemann2017}. 
This hypothesis was strongly supported by the discovery of Reticulum~II, an ultra-faint dwarf galaxy highly enhanced in heavy $r$-process elements \citep{ji2016a,roederer2016}, and recently confirmed with the detection of gravitational waves from a neutron star merger (NSM) by LIGO \citep{abbott2017}, where photometric and spectroscopic follow-up of the kilonova source SSS17a associated with GW170817 \citep{kilpatrick2017} exhibited clear evidence
for the presence of unstable isotopes created by the $r$-process
\citep{drout2017,shappee2017}.

However, it is not yet known if NSMs occur with sufficient frequency, produce sufficient amounts of $r$-process-element material, and are capable of producing the full mass range of $r$-process
elements to be the sole source of the $r$-process elements in the Universe, or whether additional sites are required \citep[see, e.g.,][]{cote2017}. Other candidates for neutron-capture-element production in the early Universe that are still not ruled out include jets in magneto-rotational supernovae \citep[Jet-SNe:][]{cameron2003,winteler2012,nishimura2015,mosta2017} and
neutrino-driven winds in core-collapse supernovae
\citep[CCSNe:][]{arcones2007,wanajo2013,thielemann2017}. CCSNe are mainly believed to contribute to the light neutron-capture elements up to $Z = 56$, whereas some Jet-SN models produce the full mass range like the NSMs \citep{nishimura2015}, while more recent models only produce the light elements \citep{mosta2017}. However, although the abundance pattern produced by the Jet-SNe and NSMs may be similar, the timescales of the two events are quite different -- massive stars explode on timescales of $\sim1-10$~Myr, while a few hundred Myr to a few Gyr are typically needed for NSMs to take place \citep{dominik2012}. 

Complementary to other efforts, progress towards solving this puzzle can be obtained by observations of very metal-poor (VMP; $\mathrm{[Fe/H]} < -2$) and extremely metal-poor (EMP; $\mathrm{[Fe/H]} < -3$) stars in the halo of the Milky Way, as such stars formed from gas polluted by one or only a few enrichment events, leaving a clear signature of the event(s) in the abundance patterns of these stars \citep{beers2005,frebel2015}. 

Different types of $r$-process element-abundance patterns exist at low metallicity. One group of stars exhibits enhancements of the heavy $r$-process elements with $Z \geq 56$. Europium (Eu; $Z = 63$) is used as a tracer of the $r$-process in these stars, as this element is almost entirely produced by the $r$-process at early times, and is the easiest of the heavy $r$-process elements to measure in optical spectra. These $r$-process-enhanced metal-poor stars are divided into two sub-classes: the moderately-enhanced $r$-I stars with $+0.3 \leq \mathrm{[Eu/Fe]} \leq +1.0$, and the highly-enhanced $r$-II stars with $\mathrm{[Eu/Fe]} > +1.0$ \citep{beers2005}. HE~1523$-$0901 \citep{frebel2007}
is an example of a typical $r$-II star; HE~0524$-$2055 \citep{barklem2005} is a typical $r$-I star. A further division of the $r$-II sub-class can be made based on the abundances of the actinide elements thorium and uranium A subset of $r$-II stars exhibit an extra enhancement in these elements, the so-called ``actinide boost" stars such as CS~31082$-$001 \citep{hill2002}. 

A second group of stars exhibits low abundances of the heavy $r$-process elements ($\mathrm{[Eu/Fe]} < +0.3$), and higher abundances of the light $r$-process elements, such as Strontium (Sr; $Z = 38$), relative to the heavy $r$-process elements ($\mathrm{[Sr/Ba]} > +0.5$). These stars display the signature of an $r$-process that is limited by a small number of neutrons, with the consequence of producing primarily light $r$-process elements (see review by \citealt{frebel2018}); we therefore label these ``limited-$r$'' stars (previously they have been described as being enriched via the ``weak" $r$-process  or Light Element Primary Process). HD~122563 is the benchmark star of this group \citep{sneden1983,honda2006}. 

Finally, a number of the $r$-process-enhanced stars also show enhancements in carbon, the so-called carbon-enhanced metal-poor (CEMP)-$r$ stars ($\mathrm{[C/Fe]} \geq +0.7$ and $\mathrm{[Eu/Fe]} > +0.3$). The first star of this type to be identified was CS~22892$-$052 \citep{sneden1994}. The original definition of these stars only includes stars with $\mathrm{[Eu/Fe]} > +1.0$ \citep{beers2005}, but we also include the $r$-I stars with carbon enhancement in this sub-class. These four sub-classes of $r$-process-enhanced stars, $r$-I, $r$-II, limited-$r$ and CEMP-$r$ (summarized in Table~\ref{tab:selection}) trace potential variations in $r$-process nucleosynthesis, as described in \citet{frebel2018}, 
the dynamics of the progenitor objects, and/or in the mixing of the $r$-process-rich ejecta with the interstellar medium, at early and late times. 

Over the past decades efforts have been made to systematically search for $r$-process-enhanced stars. \cite{christlieb2004} and \citet{barklem2005} (HERES I and II, respectively) carried out a search for $r$-II stars in the halo by taking
``snapshot" ($R\sim20,000-25,000$ and S/N$\sim$30) spectra of 253 VMP and EMP stars with VLT/UVES. They selected targets from the Hamburg/ESO Survey (HES) brighter than $B \sim 16.5$, with $0.5 < B-V < 1.2$ and $\mathrm{[Fe/H]} < -2.5$.\footnote{Note that the colors and metallicity estimates were obtained from the original survey plates, and subsequent medium-resolution spectroscopic follow-up, respectively.  As a result, the final temperature and metallicity range of the HERES sample stars were broader than originally intended.} This survey yielded 8 new $r$-II stars and 35 $r$-I stars, corresponding to frequencies of $\sim 3$\% and $\sim15$\%. A number of additional $r$-II and $r$-I stars have been added to the tally by other authors \citep[e.g.,][]{honda2004,lai2008,frebel2007,roederer2014b,mashonkina2014},
resulting in a total of $\sim30$ $r$-II stars and $\sim125$ $r$-I stars known in the halo today. For the limited-$r$ stars, a search in JINAbase for metal-poor stars shows that about 42 stars with a limited $r$-process-element signature are known today \citep{abohalima2017}. 

Additionally, spectroscopic studies of Milky Way dwarf satellite galaxies have resulted in a number of $r$-II and $r$-I stars being discovered in these systems, most recently with the discovery of the ultra-faint dwarf (UFD) galaxy Reticulum~II, where seven of nine stars analyzed exhibit large enhancements in $r$-process elements
\citep{ji2016a,ji2016b,roederer2016}. \citet{hansen2017} also reported on the discovery of an $r$-I star in the UFD galaxy Tucana~III.  More
luminous (and more massive) systems such as Ursa Minor and Draco also
contain a number of $r$-process-enhanced stars
\citep{shetrone2003,aoki2007,cohen2009,cohen2010,tsujimoto2017}. Furthermore, many metal-poor globular clusters are $r$-process enhanced and contain a large number of $r$-I stars \citep{gratton2004}. One $r$-II star with $\mathrm{[Eu/Fe]} = +1.18$ has even been detected in M15 \citep{sobeck2011}.  

Some characteristics have been found for the neutron-capture-element abundance pattern of $r$-process-enhanced stars, such as the very robust pattern seen in $r$-I and $r$-II stars from Ba to Hf, matching that of the Solar System $r$-process-abundance pattern in this range
\citep{barklem2005,siqueiramello2014}. However, the simple reality is that, once sliced into sub-samples that highlight possibly contrasting behaviors among these stars, one loses the ability to statistically quantify the results with any precision. Furthermore,
the majority of the currently known $r$-II stars are faint ($V > 13.5$), hence large amounts of telescope time are needed to obtain spectra with sufficient S/N to derive abundances for a large number of neutron-capture elements in these stars.

To further advance our understanding of the $r$-process, we have established a collaboration known as the $R$-Process Alliance (RPA), which will combine observations, theory and modeling, and experiments from multiple fields to investigate different aspects of the $r$-process. The first goal in this effort is to carry out an extensive search for $r$-process-enhanced stars in our Galaxy, in order to increase the number of known $r$-II and limited-$r$ stars from $\sim 30$ and $\sim 40$, respectively, to $\sim 100-125$ each (we expect some 500 $r$-I stars to be discovered in the process as well), and thus compile a much larger ``statistical" sample of such stars that is sufficiently bright to facilitate abundance determinations for a large number of neutron-capture elements. 

This paper describes the first data sample from the Southern Hemisphere search. A second paper in this series (Sakari et al. 2018b, in prep.) will present the first data sample from the Northern Hemisphere search. Another paper reporting on more Southern Hemisphere targets is  currently being prepared. From this large-scale search effort, we have already found several interesting $r$-process-enhanced stars (see Section 2.2), as well as the first bona-fide CEMP-$r+s$ star \citep{gull2018}.

\begin{deluxetable}{lc}[t]
\tablecaption{Abundance Signatures for $r$-II, $r$-I, limited-$r$, and
  CEMP-$r$ stars \label{tab:selection}}
\tablewidth{0pt}
\tablehead{ \colhead{Sub-class} & \colhead{Abundances} }
\startdata 
$r$-II   & $\mathrm{[Eu/Fe]} > +1.0$, $\mathrm{[Ba/Eu]} < 0$ \\
$r$-I    & $+0.3 \leq \mathrm{[Eu/Fe]} \leq +1.0$, $\mathrm{[Ba/Eu]} < 0$ \\
limited-$r$& $\mathrm{[Sr/Ba]} > +0.5$, $\mathrm{[Eu/Fe]} < +0.3$\\
CEMP-$r$\tablenotemark{a} & $\mathrm{[C/Fe]} \geq +0.7$, $\mathrm{[Eu/Fe]} > +0.3$, $\mathrm{[Ba/Eu]} < 0$ \\
\enddata
\tablenotetext{a}{The original definition of these stars only includes stars with $\mathrm{[Eu/Fe]} > +1.0$ \citep{beers2005}, but we also include the $r$-I stars with carbon enhancement in this sub-class.}
\end{deluxetable}  

\section{Observations}
\label{observations}

\subsection{Strategy}
\label{sec:strategy} 
From the survey carried out by \citet{barklem2005}, the expected frequencies of $r$-II and  $r$-I stars in the halo system are $\sim3$\% and $\sim$15\%, respectively. Among the 1658 metal-poor stars in JINAbase, 42 stars have abundances that match our limited-$r$ definition, resulting in a frequency of $\sim3$\% for this sub-class
\citep{abohalima2017}. This frequency of limited-$r$ stars is a lower limit, as many candidates with high upper limits on their Eu abundance exist, and these could belong to the limited-$r$ (or $r$-I) sub-classes. However, it is clear that only a few stars belonging to this sub-class have data available to facilitate abundance derivation for a large number of heavy neutron-capture elements. In order to significantly increase the number of known $r$-II and limited-$r$ stars in the halo we need to obtain snapshot spectra for large numbers of bright stars, which facilitate abundance determination for a large number of neutron-capture elements.  

To achieve the first goal of the RPA, we plan to obtain snapshot spectra of $\sim$2500 stars with $V \lesssim 13.5$ and $\mathrm{[Fe/H]} \le -2.0$, over both the Southern and Northern Hemispheres, using a variety of moderate- to large-aperture telescopes. Here we present the first sample, taken with the Echelle spectrograph at the Las Campanas Observatory du Pont 2.5m telescope. In a second paper, we report on a similar effort carried out with the Apache Point Observatory 3.5m telescope (Sakari et al. 2018b, in prep.), with additional papers to follow.

Although several tens of thousands of candidate stars with $\mathrm{[Fe/H]} < -2$ are now known (primarily from SDSS/SEGUE; \citealt{abazajian2009,yanny2009}), the great majority are too faint to efficiently obtain high-resolution data for a large number of stars, in order to enable the determination of accurate abundances.  
Fortunately, a number of recent large surveys have provided suitably bright stars for consideration. Unfortunately, not all of these stars were known a-priori to be VMP or EMP stars, thus Phase I of the RPA search for $r$-process-enhanced stars was to obtain medium-resolution validation data for those targets. These include stars from the RAVE survey (DR4; \citealt{kordopatis2013}, DR5; \citealt{kunder2017}) and the Best \& Brightest Survey \citep[][B\&B]{schlaufman2014}. Note that, although the RAVE data releases provided estimates of metallicity for many bright stars, we have found that only about 60\% of the stars from this survey with quoted $\mathrm{[Fe/H]} < -2$ are in fact this metal poor \citep{placco2018}. We therefore decided to validate as many of these stars as possible before proceeding to the high-resolution observations. We have also included stars satisfying our selection criteria that already have had medium-resolution validation completed. These include stars from the HK Survey of Beers and collaborators \citep[][and Beers et al., in  prep.]{beers1985,beers1992} or the Hamburg/ESO Survey of Christlieb and colleagues \citep[][Beers et al., in prep.]{christlieb2008,frebel2006,beers2017}, from the SkyMapper
Survey \citep{wolf2018}, and (in particular in the Northern Hemisphere) G-K-giants selected from the LAMOST survey (Beers et al., in prep.) Stars from smaller lists of candidates from various programs carried out by members of our team were also included. 

For Phase II, we selected targets with $V < 13.5$,  $\mathrm{[Fe/H]} < -2$, 4000~K $< T_{\rm eff} <$ 5500~K, and not strongly enhanced in carbon, as validated by either extant medium-resolution spectra in our possession or spectra that we gathered as part of Phase I of the search  
(see \citet{placco2018}, for a more detailed description of our
target selection and determination of atmospheric parameters and
$\mathrm{[C/Fe]}$ using the n-SSPP, described by \citealt{beers2014}).  
We then obtained snapshot high-resolution spectra with a S/N of $\sim$ 30 at 4100\,{\AA} and a resolving power $R \sim$ 25,000. This can be achieved in a 1200 sec exposure for a $V=11.5$ star with a 2.5m-class
telescope. \citet{christlieb2004} and \citet{barklem2005} established that such spectra are sufficient to detect (or derive meaningful upper limits on) an enhanced Eu line at 4129\,{\AA} and a Sr line at 4077\ {\AA}, for cool, VMP stars. We then derive C, Sr, Ba, and Eu abundances for these stars, which allow us to identify likely $r$-II, $r$-I, limited-$r$, and CEMP-$r$ stars, following the selection criteria summarized in Table \ref{tab:selection}. When identifying $r$-II, $r$-I, and CEMP-$r$ stars from spectra taken in Phase II, we also
require that our selected stars satisfy $\mathrm{[Ba/Eu]} < 0$, to ensure that the chemical composition of the stars is dominated by the $r$-process and not the slow neutron-capture process ($s$-process; \citealt{beers2005}).  

Phase III of this effort, which is already underway, makes use of generally large-aperture telescopes in order to obtain much higher-quality ``portrait" spectra (S/N $\sim 100$ at 3850\,{\AA}; $R \sim 50,000$ or higher) for all of the $r$-II and limited-$r$ stars (and a subset of the $r$-I stars) identified during Phase II. \citet{placco2017} reports on the first portrait spectrum taken during Phase III, for RAVE~J203843.2$-$002333, a bright ($V = 12.7$), VMP ($\mathrm{[Fe/H]} = -2.91$), $r$-II ($\mathrm{[Eu/Fe]} = +1.64$ and $\mathrm{[Ba/Eu]} = -0.81$) star originally identified during the RAVE survey; it is also only the fourth VMP/EMP star reported with a measured uranium abundance. In addition, RAVE~J153830.9$-$180424, one of the most metal-rich $r$-II stars known in the Milky Way halo ($\mathrm{[Fe/H]}= -2.09$), was recently identified in the Northern Hemisphere Phase-II search. The portrait spectrum of this star, and its elemental-abundance analysis, is presented in \cite{sakari2018}. The full set of portrait spectra will be analyzed and published in due course.

\subsection{The Southern Hemisphere Search for $r$-Process-Enhanced Stars}

The first dataset on this program in the Southern Hemisphere was obtained during six nights in August 2016, with the Echelle spectrograph on the du Pont 2.5m telescope at the Las Campanas Observatory. During this run, we obtained spectra for 107 stars in the magnitude range 10 $<$ V $<$ 13.5 and with metallicities of $-3.13 < \mathrm{[Fe/H]} < -0.79$ (note that these metallicities reflect the high-resolution abundance results, not those from Phase I). The spectra were obtained using the 1{\arcsec}x4{\arcsec} slit and 2x2 on-chip binning, yielding a resolving power of $R\sim25,000$, covering the wavelength range from  3860\,{\AA} to 9000\,{\AA}. The data were reduced using
the Carnegie Python Distribution\footnote{http://code.obs.carnegiescience.edu/}
\citep{kelson1998,kelson2000,kelson2003}. Radial velocities were measured from order-by-order cross-correlation of the target spectra with a spectrum of HD~213575, a radial-velocity standard star obtained during the run, using the \emph{fxcor} task in IRAF \footnote{IRAF is distributed by the National Optical Astronomy Observatory, which is operated by the Association of Universities for Research in Astronomy, Inc., under cooperative agreement with the NSF.}.  
On average 17 orders in each spectrum were used for the correlation. We also estimated the S/N of each spectrum in the region of the 4129\,{\AA} Eu line by taking the square root of the total counts in the continuum, converted to photons by multiplying with the CCD gain =1.05 e$^{-}$/DN. 
Table~\ref{tab:obslog} lists the observed targets with RA, DEC, $V$
magnitude, MJD, exposure time, S/N at 4129\,{\AA}, heliocentric radial
velocity (RV$_{\rm helio}$) and its associated error, the source catalog from which we originally selected each target, and additional common identifiers for the stars.  

\begin{longrotatetable}[t!]
\begin{deluxetable*}{lrrrrrrrrll}
\tabletypesize{\scriptsize}
\tablecaption{Observation Log \label{tab:obslog}}
\tablewidth{0pt}
\tablehead{
\colhead{Stellar ID} & \colhead{RA} & \colhead{Dec} & \colhead{$V$ mag\tablenotemark{a}}
&\colhead{MJD} &\colhead{Exp Time} & \colhead{S/N}  & \colhead{RV$_{\rm helio}$}
& \colhead{RV$_{\rm err}$} &\colhead{Source} &\colhead{Other Identifier} \\
\colhead{} & \colhead{} & \colhead{} & \colhead{}
&\colhead{} &\colhead{(s)} & \colhead{4129\,{\AA} } & \colhead{(km~s$^{-1}$)}
& \colhead{(km~s$^{-1}$)}&\colhead{}&\colhead{}}
\startdata 
J00002259$-$1302275 &  00 00 22.6  &  $-$13 02 27.5  &  12.9  &  57605.82086  &  296   &  17  &  $-$102.22  &  0.75  &  RAVE DR5 \\
  		    &    &    &    &  57607.81559  &  900  &  31  &  $-$102.24  &  1.25  &  \\
J00021222$-$2241388 &  00 02 12.2  &  $-$22 41 38.9  &  13.3  &  57606.83137  &  3600  &  24  &  $+$35.24  &  0.76  &  B\&B \\
J00021668$-$2453494 &  00 02 16.7  &  $-$24 53 49.5  &  13.5  &  57607.83983  &  3600  &  31  &  $+$78.82  &  0.49  &  B\&B \\
J00133067$-$1259594 &  00 13 30.7  &  $-$12 59 59.4  &  11.8  &  57608.80407  &  1000  &  31  &  $-$95.29  &  0.31  &  RAVE DR5 & HE~0010$-$1316\\
J00233067$-$1631428 &  00 23 30.7  &  $-$16 31 43.2  &  12.3  &  57609.83271  &  1800  &  30  &  $-$8.78   &  0.55  &  RAVE DR5 \\
J00400685$-$4325183 &  00 40 06.9  &  $-$43 25 18.4  &  12.4  &  57606.86896  &  600   &  20  &  $+$48.93  &  0.74  &  RAVE DR5 & HE~0037$-$4341\\
J00405260$-$5122491 &  00 40 52.6  &  $-$51 22 49.1  &  11.2  &  57609.81417  &  900   &  34  &  $+$123.11 &  0.21  &  RAVE DR5 & CD$-$52~112\\
J00453930$-$7457294 &  00 45 39.3  &  $-$74 57 29.6  &  11.7  &  57608.82169  &  1200  &  31  &  $-$31.23  &  0.41  &  RAVE DR5 \\
J01202234$-$5425582 &  01 20 22.3  &  $-$54 25 58.2  &  12.6  &  57606.89108  &  2700  &  32  &  $+$68.00  &  0.85  &  RAVE DR5 \\
J01293113$-$1600454 &  01 29 31.1  &  $-$16 00 45.5  &  11.6  &  57609.84994  &  800   &  25  &  $+$139.47 &  0.30  &  RAVE DR5& CS~31082-001, BD$-$16~251 \\
J01334657$-$2727374 &  01 33 46.6  &  $-$27 27 37.3  &  12.4  &  57608.85405  &  3600  &  42  &  $-$25.87  &  0.61  &  RAVE DR5 \\
J01425422$-$5032488 &  01 42 54.2  &  $-$50 32 48.9  &  11.4  &  57609.86360  &  1000  &  33  &  $+$158.15 &  0.17  &  RAVE DR5& \\
J01430726$-$6445174 &  01 43 07.3  &  $-$64 45 17.4  &  12.0  &  57610.86966  &  2100  &  35  &  $+$167.91 &  0.24  &  RAVE DR5& \\
J01451951$-$2800583 &  01 45 19.5  &  $-$28 00 58.4  &  11.7  &  57608.88461  &  1100  &  38  &  $+$34.89  &  0.28  &  RAVE DR5& \\
J01490794$-$4911429 &  01 49 07.9  &  $-$49 11 43.0  &  11.2  &  57610.80049  &  1200  &  42  &  $+$90.30  &  0.25  &  RAVE DR5& HE~0147$-$4926, CD-49~506 \\
J01530024$-$3417360 &  01 53 00.2  &  $-$34 17 36.1  &  9.6   &  57606.91442  &  900   &  72  &  $+$22.81  &  0.33  &  RAVE DR5& HD~11582, CD$-$34~722  \\
J02040793$-$3127556 &  02 04 07.9  &  $-$31 27 55.6  &  12.6  &  57607.90877  &  1900  &  37  &  $+$83.45  &  1.00  &  RAVE DR5& HE~0201$-$3142 \\
J02165716$-$7547064 &  02 16 57.2  &  $-$75 47 06.6  &  11.9  &  57608.90250  &  1200  &  31  &  $-$5.80   &  0.27  &  RAVE DR5 \\
J02355867$-$6745520 &  02 35 58.7  &  $-$67 45 51.9  &  11.9  &  57610.82171  &  1800  &  23  &  $+$101.45 &  0.19  &  RAVE DR5 \\
J02401075$-$1416290 &  02 40 10.8  &  $-$14 16 29.9  &  13.1  &  57606.92573  &  600   &  19  &  $+$126.28 &  1.80  &  RAVE DR5 \\
J02412152$-$1825376 &  02 41 21.5  &  $-$18 25 37.8  &  11.7  &  57608.92031  &  1200  &  34  &  $-$108.51 &  0.24  &  RAVE DR5& \\
J02441479$-$5158241 &  02 44 14.8  &  $-$51 58 24.1  &  11.8  &  57609.88054  &  1400  &  32  &  $+$52.13  &  0.36  &  RAVE DR5& HE~0242$-$5211\\
J02462013$-$1518419 &  02 46 20.1  &  $-$15 18 41.9  &  12.5  &  57609.92227  &  600   &  20  &  $+$278.23 &  0.39  &  RAVE DR5& SDSS~J024620.14-151842.0 \\
J02500719$-$5145148 &  02 50 07.2  &  $-$51 45 15.1  &  12.4  &  57610.90069  &  2700  &  32  &  $+$310.48 &  0.22  &  RAVE DR5 \\
J03193531$-$3250433 &  03 19 35.3  &  $-$32 50 43.3  &  10.9  &  57610.84230  &  1200  &  52  &  $+$47.95  &  1.52  &  RAVE DR5& HE~0317$-$3301, CD-33~1173 \\
J03563703$-$5838281 &  03 56 37.0  &  $-$58 38 28.1  &  11.4  &  57609.89703  &  900   &  22  &  $+$157.07 &  0.74  &  RAVE DR5& HE~0355$-$5847 \\
J04121388$-$1205050 &  04 12 13.9  &  $-$12 05 05.1  &  12.9  &  57609.91101  &  900   &  19  &  $+$14.40  &  0.96  &  RAVE DR5& HE~0409$-$1212, CS~22169-0035 \\
J13164824$-$2743351 &  13 16 48.2  &  $-$27 43 35.2  &  11.5  &  57605.50624  &  1200  &  32  &  $+$69.00  &  0.19  &  RAVE DR5& \\
J14033542$-$3335257 &  14 03 35.4  &  $-$33 35 25.7  &  11.5  &  57609.46903  &  600   &  29  &  $+$262.39 &  0.19  &  RAVE DR5& CD$-$33~9514  \\
J14164084$-$2422000 &  14 16 40.8  &  $-$24 21 59.9  &  10.9  &  57609.47870  &  600   &  30  &  $+$0.84   &  0.26  &  RAVE DR5& \\
J14232679$-$2834200 &  14 23 26.8  &  $-$28 34 20.1  &  12.7  &  57607.48495  &  2700  &  33  &  $+$134.07 &  0.50  &  RAVE DR5 \\
J14301385$-$2317388 &  14 30 13.9  &  $-$23 17 38.7  &  12.0  &  57608.49486  &  400   &  25  &  $+$432.11 &  0.29  &  RAVE DR5 \\
J14325334$-$4125494 &  14 32 53.3  &  $-$41 25 49.5  &  11.1  &  57605.49093  &  1200  &  36  &  $-$227.92 &  0.14  &  B\&B&  \\
J14381119$-$2120085 &  14 38 11.2  &  $-$21 20 08.7  &  11.4  &  57609.49022  &  1000  &  26  &  $+$9.22   &  0.19  &  RAVE DR5 \\
J15260106$-$0911388 &  15 26 01.1  &  $-$09 11 38.7  &  11.1  &  57609.51084  &  400   &  22  &  $-$162.27 &  0.41  &  RAVE DR5& HE~1523$-$0901  \\
J15271353$-$2336177 &  15 27 13.5  &  $-$23 36 17.7  &  10.8  &  57609.52091  &  900   &  45  &  $+$1.22   &  0.23  &  RAVE DR5& CD$-$23~12296 \\
J15575183$-$0839549 &  15 57 51.8  &  $-$08 39 55.0  &  10.4  &  57609.53186  &  600   &  28  &  $-$21.32  &  0.15  &  RAVE DR5& BD-08~4115 \\
J15582962$-$1224344 &  15 58 29.6  &  $-$12 24 34.4  &  12.3  &  57605.52399  &  1200  &  28  &  $+$83.08  &  0.42  &  RAVE DR5 \\
J16024498$-$1521016 &  16 02 45.0  &  $-$15 21 01.7  &  10.3  &  57606.55986  &  900   &  34  &  $+$101.26 &  0.45  &  RAVE DR5 & BD$-$14~4336  \\
J16080681$-$3215592 &  16 08 06.8  &  $-$32 15 59.3  &  12.7  &  57610.57131  &  2200  &  29  &  $+$163.49 &  0.41  &  B\&B \\
J16095117$-$0941174 &  16 09 51.2  &  $-$09 41 17.5  &  10.5  &  57608.50559  &  1000  &  53  &  $+$129.61 &  0.15  &  RAVE DR5& BD$-$09~4312  \\
J16103106$+$1003055 &  16 10 31.1  &  $+$10 03 05.6  &  12.8  &  57607.52596  &  2400  &  33  &  $-$68.33  &  0.95  &  B\&B \\
J16184302$-$0630558 &  16 18 43.1  &  $-$06 30 55.5  &  12.1  &  57610.54438  &  1800  &  33  &  $+$80.91  &  0.43  &  RAVE DR5 \\
J17093199$-$6027271 &  17 09 32.0  &  $-$60 27 27.1  &  12.5  &  57605.54109  &  600   &  19  &  $+$332.85 &  0.37  &  RAVE DR5 \\
   		    &    &    &    &  57608.51985  &  600  &  25  &  $+$333.73  &  0.38  &  \\
J17094926$-$6239285 &  17 09 49.3  &  $-$62 39 28.6  &  12.6  &  57610.51453  &  2700  &  27  &  $-$50.98  &  0.30  &  RAVE DR5 \\
J17124284$-$5211479 &  17 12 42.8  &  $-$52 11 47.9  &  13.1  &  57608.53750  &  1800  &  29  &  $-$26.56  &  1.28  &  B\&B \\
J17225742$-$7123000 &  17 22 57.4  &  $-$71 22 60.0  &  12.9  &  57605.55568  &  1200  &  21  &  $+$277.27 &  0.51  &  RAVE DR5 \\
		    &    &    &    &  57607.54850  &  600  &  24  &  $+$278.07  &  1.08  &  \\
J17255680$-$6603395 &  17 25 56.8  &  $-$66 03 39.6  &  12.7  &  57609.55763  &  2700  &  27  &  $+$355.68 &  0.60  &  B\&B \\
J17273886$-$5133298 &  17 27 38.9  &  $-$51 33 30.0  &  12.1  &  57608.59039  &  1500  &  34  &  $-$219.14 &  0.41  &  B\&B \\
J17285930$-$7427532 &  17 28 59.3  &  $-$74 27 53.2  &  13.1  &  57607.57677  &  3600  &  33  &  $+$260.49 &  0.79  &  B\&B \\
J17334285$-$5121550 &  17 33 42.9  &  $-$51 21 55.1  &  12.5  &  57608.55990  &  1500  &  32  &  $+$105.35 &  0.24  &  B\&B \\
J17400682$-$6102129 &  17 40 06.8  &  $-$61 02 12.9  &  10.3  &  57606.57576  &  1200  &  43  &  $+$255.02 &  0.28  &  RAVE DR5& CD$-$60~6745 \\
J17435113$-$5359333 &  17 43 51.1  &  $-$53 59 33.4  &  12.0  &  57605.57992  &  1000  &  28  &  $+$22.53  &  0.23  &  RAVE DR5 \\
J18024226$-$4404426 &  18 02 42.3  &  $-$44 04 42.6  &  12.6  &  57605.56963  &  300   &  18  &  $-$226.40 &  0.32  &  RAVE DR5 \\
  		    &    &    &    &  57608.57499  &  500  &  26  &  $-$226.06  &  1.72  &  \\
J18121045$-$4934495 &  18 12 10.5  &  $-$49 34 49.5  &  12.6  &  57608.61337  &  1800  &  34  &  $+$36.61  &  0.37  &  B\&B \\
J18285086$-$3434203 &  18 28 50.9  &  $-$34 34 20.4  &  12.7  &  57605.60844  &  1600  &  28  &  $+$22.73  &  0.33  &  B\&B \\
J18294122$-$4504000 &  18 29 41.2  &  $-$45 04 00.1  &  12.9  &  57606.59626  &  3600  &  22  &  $-$108.28 &  1.23  &  B\&B \\
J18362318$-$6428124 &  18 36 23.2  &  $-$64 28 12.5  &  12.5  &  57609.59046  &  2400  &  29  &  $+$16.21  &  0.32  &  B\&B \\
J18363613$-$7136597 &  18 36 36.1  &  $-$71 36 59.8  &  10.9  &  57605.59266  &  500   &  23  &  $-$68.32  &  0.20  &  RAVE DR5& \\
J18562774$-$7251331 &  18 56 27.7  &  $-$72 51 33.1  &  12.7  &  57610.60588  &  2800  &  26  &  $+$177.40 &  0.52  &  B\&B \\
J19014952$-$4844359 &  19 01 49.5  &  $-$48 44 35.9  &  12.5  &  57607.61450  &  2400  &  38  &  $+$133.36 &  0.32  &  B\&B \\
J19161821$-$5544454 &  19 16 18.2  &  $-$55 44 45.4  &  11.5  &  57605.62378  &  450   &  22  &  $+$46.58  &  0.29  &  RAVE DR5& \\
J19172402$-$4211240 &  19 17 24.0  &  $-$42 11 24.1  &  12.8  &  57605.64483  &  2700  &  34  &  $+$37.44  &  0.42  &  RAVE DR5 \\
J19173310$-$6628544 &  19 17 33.1  &  $-$66 28 54.5  &  13.1  &  57609.61946  &  4200  &  21  &  $+$62.18  &  0.52  &  B\&B \\
J19202070$-$6627202 &  19 20 20.7  &  $-$66 27 20.3  &  12.6  &  57607.64836  &  2700  &  35  &  $+$6.73   &  0.33  &  B\&B \\
J19215077$-$4452545 &  19 21 50.8  &  $-$44 52 54.6  &  11.4  &  57606.65080  &  3000  &  31  &  $+$3.53   &  0.50  &  RAVE DR5 & \\
J19232518$-$5833410 &  19 23 25.2  &  $-$58 33 40.9  &  11.9  &  57606.63524  &  700   &  23  &  $+$125.88 &  0.57  &  RAVE DR5 &  \\
J19324858$-$5908019 &  19 32 48.6  &  $-$59 08 01.9  &  12.8  &  57608.63256  &  700   &  24  &  $+$303.44 &  0.50  &  RAVE DR5 \\
J19345497$-$5751400 &  19 34 55.0  &  $-$57 51 40.1  &  13.1  &  57608.64767  &  1300  &  27  &  $-$81.91  &  0.55  &  RAVE DR5 \\
J19494025$-$5424113 &  19 49 40.3  &  $-$54 24 11.2  &  12.7  &  57610.64351  &  3000  &  33  &  $+$61.07  &  0.49  &  RAVE DR5 \\
J19534978$-$5940001 &  19 53 49.8  &  $-$59 40 00.1  &  12.6  &  57608.66299  &  600   &  27  &  $+$216.45 &  0.55  &  RAVE DR5& CS~22873-0055 \\
J19594558$-$2549075 &  19 59 45.5  &  $-$25 49 07.0  &  12.2  &  57609.50547  &  1200  &  22  &  $-$205.59 &  0.80  &  RAVE DR5 \\
J20093393$-$3410273 &  20 09 33.9  &  $-$34 10 27.3  &  11.8  &  57607.68078  &  2400  &  38  &  $+$27.24  &  0.30  &  RAVE DR5 \\
J20144608$-$5635300 &  20 14 46.1  &  $-$56 35 30.1  &  12.2  &  57605.67459  &  1200  &  23  &  $-$11.49  &  0.39  &  RAVE DR5 \\
J20165358$-$0503593 &  20 16 53.6  &  $-$05 03 59.3  &  12.5  &  57608.68904  &  2000  &  33  &  $+$27.34  &  0.72  &  B\&B \\
J20303339$-$2519500 &  20 30 33.4  &  $-$25 19 50.2  &  10.4  &  57605.68612  &  500   &  34  &  $-$57.98  &  0.13  &  RAVE DR5& CD$-$25~14822 \\
J20313531$-$3127319 &  20 31 35.3  &  $-$31 27 31.9  &  13.6  &  57609.71514  &  1200  &  19  &  $-$223.19 &  2.15  &  RAVE DR5 \\
 		    &    &    &    &  57610.68591  &  3600  &  33  &  $-$222.53  &  0.16  &  \\
J20445584$-$2250597 &  20 44 55.9  &  $-$22 50 59.8  &  11.3  &  57608.71550  &  600   &  33  &  $+$51.86  &  0.16  &  RAVE DR5& \\
J20492765$-$5124440 &  20 49 27.7  &  $-$51 24 44.0  &  11.5  &  57605.69543  &  600   &  25  &  $+$24.56  &  0.20  &  RAVE DR5 \\
J20514971$-$6158008 &  20 51 49.7  &  $-$61 58 00.9  &  12.3  &  57609.67235  &  1500  &  27  &  $-$93.62  &  0.40  &  RAVE DR5 \\
J20542346$-$0033097 &  20 54 23.5  &  $-$00 33 09.8  &  11.7  &  57610.71914  &  1500  &  29  &  $-$61.92  &  0.25  &  RAVE DR5&  \\
J20560913$-$1331176 &  20 56 09.1  &  $-$13 31 17.7  &  10.2  &  57607.70042  &  500   &  51  &  $+$123.45 &  0.11  &  RAVE DR5& HD~358059, BD$-$14~5890 \\
J21023752$-$5316132 &  21 02 37.5  &  $-$53 16 13.3  &  12.9  &  57605.71525  &  2400  &  30  &  $-$3.34   &  0.48  &  B\&B \\
J21063474$-$4957500 &  21 06 34.7  &  $-$49 57 50.1  &  9.0   &  57608.72475  &  400   &  80  &  $-$44.13  &  0.29  &  RAVE DR5& HD~200654, CD$-$50~13237 \\
J21064294$-$6828266 &  21 06 42.9  &  $-$68 28 26.7  &  12.8  &  57608.73613  &  900   &  27  &  $-$72.65  &  0.58  &  RAVE DR5 \\
J21091825$-$1310066 &  21 09 18.2  &  $-$13 10 06.2  &  10.7  &  57606.70093  &  550   &  32  &  $-$35.91  &  0.39  &  RAVE DR5& HD~358246 \\
J21095804$-$0945400 &  21 09 58.1  &  $-$09 45 40.0  &  13.2  &  57606.71019  &  700   &  17  &  $-$20.59  &  1.58  &  RAVE DR5 \\
		    &    &    &    &  57607.71303  &  700  &  24  &  $-$17.39  &  1.74  &  \\
   		    &    &    &    &  57608.75826  &  2200  &  27  &  $-$17.20  &  1.97  &  \\
J21141350$-$5726363 &  21 14 13.5  &  $-$57 26 36.2  &  10.5  &  57607.72478  &  700   &  51  &  $+$158.37  &  0.08  &  RAVE DR5& V*~AC~Aqr, BD$-$02~5494  \\
J21152551$-$1503309 &  21 15 25.5  &  $-$15 03 31.0  &  12.6  &  57609.74547  &  2400  &  22  &  $-$6.05    &  0.78  &  RAVE DR5& HE~2119$-$4653 \\
J21162185$-$0213420 &  21 16 21.9  &  $-$02 13 42.0  &  10.2  &  57609.72698  &  400   &  21  &  $-$5.32    &  0.70  &  RAVE DR5 \\
J21224590$-$4641030 &  21 22 45.9  &  $-$46 41 03.1  &  12.3  &  57609.76828  &  1000  &  21  &  $-$100.75  &  0.45  &  RAVE DR5 \\
J21262525$-$2144243 &  21 26 25.3  &  $-$21 44 24.3  &  12.1  &  57606.73219  &  2700  &  32  &  $-$81.73   &  0.62  &  RAVE DR5 \\
J21291592$-$4236219 &  21 29 15.9  &  $-$42 36 22.0  &  13.0  &  57607.74035  &  1500  &  31  &  $-$143.15  &  0.43  &  RAVE DR5 \\
J21370807$-$0927347 &  21 37 08.1  &  $-$09 27 34.6  &  13.3  &  57605.74090  &  1500  &  21  &  $-$63.81   &  0.39  &  RAVE DR5 \\
J21513595$-$0543398 &  21 51 36.0  &  $-$05 43 39.8  &  10.0  &  57609.78191  &  800   &  35  &  $-$209.08  &  0.15  &  RAVE DR5& HD~207785, BD$-$06~5849 \\
J22021636$-$0536483 &  22 02 16.4  &  $-$05 36 48.4  &  12.4  &  57609.79579  &  1200  &  21  &  $-$106.65  &  0.88  &  RAVE DR5& HE~2159$-$0551 \\
J22163596$+$0246171 &  22 16 36.0  &  $+$02 46 17.1  &  12.7  &  57605.76641  &  1900  &  28  &  $-$98.32   &  0.41  &  B\&B \\
J22444701$-$5617540 &  22 44 47.0  &  $-$56 17 54.1  &  12.5  &  57606.75810  &  1200  &  24  &  $-$189.75  &  0.92  &  RAVE DR5 \\
J22492756$-$2238289 &  22 49 27.6  &  $-$22 38 29.0  &  12.0  &  57606.77131  &  1200  &  27  &  $-$1.42    &  0.43  &  RAVE DR5&   \\
J22531984$-$2248555 &  22 53 19.9  &  $-$22 48 55.5  &  11.1  &  57607.78387  &  3600  &  71  &  $-$6.97    &  0.14  &  RAVE DR5& CD$-$23~17654  \\
J22595884$-$1554182 &  22 59 58.8  &  $-$15 54 18.2  &  10.9  &  57606.79521  &  600   &  30  &  $+$19.37   &  0.51  &  RAVE DR5& BD$-$16~6190 \\
J23022289$-$6833233 &  23 02 22.9  &  $-$68 33 23.3  &  11.6  &  57608.78529  &  1500  &  37  &  $+$136.05  &  0.23  &  RAVE DR5& \\
J23044914$-$4243477 &  23 04 49.1  &  $-$42 43 47.8  &  12.1  &  57605.79075  &  400   &  21  &  $-$278.06  &  0.50  &  RAVE DR5 & HE~2302$-$4259  \\
J23130003$-$4507066 &  23 13 00.0  &  $-$45 07 06.6  &  11.2  &  57610.78319  &  1200  &  39  &  $-$94.59   &  0.19  &  RAVE DR5& \\
J23265258$-$0159248 &  23 26 52.6  &  $-$01 59 24.8  &  11.2  &  57605.80255  &  1000  &  33  &  $-$215.33  &  0.31  &  RAVE DR5& HE~2324$-$0215, BD$-$02~5957 \\
J23310716$-$0223301 &  23 31 07.2  &  $-$02 23 30.1  &  11.5  &  57606.80969  &  1500  &  29  &  $-$99.40   &  0.77  &  RAVE DR5&  \\
J23362202$-$5607498 &  23 36 22.0  &  $-$56 07 49.8  &  12.9  &  57605.81383  &  300   &  17  &  $+$207.95  &  0.35  &  RAVE DR5 \\
  		    &    &    &    &  57607.81017  &  300  &  23  &  $+$209.55  &  0.53  &  \\
\enddata 
\tablenotetext{a}{RAVE DR5 $V$ magnitudes are from \citet{munari2014}, B\&B are from \citet{henden2014}}
\end{deluxetable*}
\end{longrotatetable}

\section{Stellar Parameter Derivations and Abundance Analysis}
\label{analysis}

\subsection{Parameters}
\begin{figure}[t!]
\centering
\includegraphics[scale=0.5]{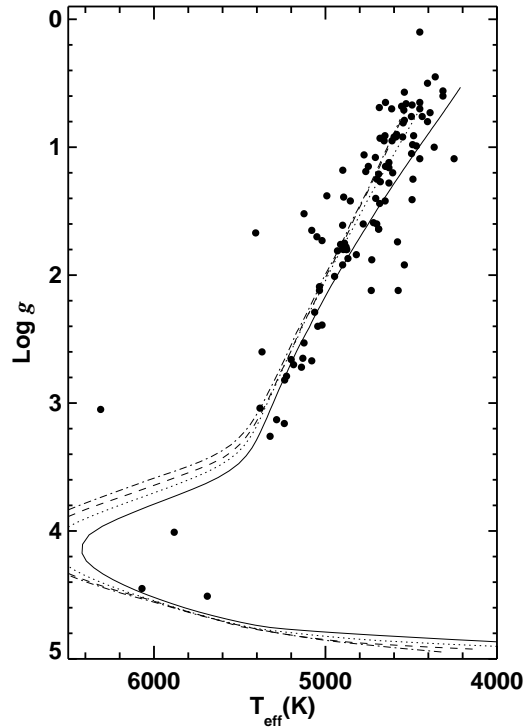}
\caption{Surface gravity versus effective temperature measurements for the
  sample stars, over-plotted with Y$^2$ isochrones for four different
  metallicities, solid: $\mathrm{[Fe/H]=-2.0}$, dotted:
  $\mathrm{[Fe/H]=-2.5}$, dashed: $\mathrm{[Fe/H]=-3.0}$, dot-dashed:
  $\mathrm{[Fe/H]=-3.5}$. \label{fig:Tefflogg}} 
\end{figure}

We derive 1D LTE stellar parameters for our program stars from 
equivalent-width measurements of a large number of \ion{Fe}{1} and \ion{Fe}{2} lines, using the 2017 version of MOOG \citep{sneden1973}, including Rayleigh scattering treatment as described by \citet{sobeck2011}\footnote{https://github.com/alexji/moog17scat}. The number of Fe lines used for the analysis ranges from 28 to 140 for \ion{Fe}{1} and 6 to 27 for \ion{Fe}{2}, with a mean of 86 and 15 lines used per spectrum, respectively. Equivalent widths of \ion{Fe}{1} and \ion{Fe}{2} lines measured for the individual stars, along with the wavelength ($\lambda$), excitation potential ($\chi$), oscillator strength ($\log gf$), and derived abundances ($\log\epsilon$) for the lines, are listed in Table \ref{tab:ew}. 
Effective temperatures ($T_{\rm eff}$) were derived by ensuring excitation equilibrium for the \ion{Fe}{1}-line abundances. Spectroscopic derivation of stellar parameters using non-local thermodynamic equilibrium (NLTE) corrected Fe abundances has been found to agree better with photometric temperatures than what is found from LTE spectroscopic temperatures \citep{ezzeddine2017}. We therefore place these initial temperatures on a photometric scale, correcting for the offset between spectroscopic and photometric temperature scales using the following relation from \citet{frebel2013}:
	\[T_{\rm{eff, corrected}} = T_{\rm{eff, initial}} -0.1\times T_{\rm{eff, initial}} + 670.
	\] 
The shift in temperature varies from 25~K for the hottest stars to 265~K for the coolest stars, with a mean shift of 208~K, owing to the predominantly low-temperature range of the sample. Surface gravities ($\log g$) were derived through ionization equilibration by ensuring  agreement between abundances derived from \ion{Fe}{1} and \ion{Fe}{2} lines. Microturbulent velocities ($\xi$) were determined by removing any trend in line abundances with reduced equivalent widths for both \ion{Fe}{1} and \ion{Fe}{2} lines. Final parameters are listed
in Table \ref{tab:param}. We adopt an uncertainty of 150~K on $T_{\rm eff}$, which reflects the uncertainty in the spectroscopically-determined $T_{\rm eff}$ and a possible scale error from the application of the above relation. The standard deviation of the derived \ion{Fe}{1} abundances varies from 0.09~dex to 0.28~dex, with a mean of 0.16~dex. Based on this value and the uncertainty in $T_{\rm eff}$, we estimate uncertainties of 0.3~dex for $\log g$ and 0.3 km~s$^{-1}$ for $\xi$. Figure \ref{fig:Tefflogg} shows the derived gravities as a function of effective temperature for the sample stars. Over-plotted are four Y$^2$ isochrones with $\mathrm{[\alpha/Fe]} = +0.4$, age 12~Gyr, and $\mathrm{[Fe/H]}= -2.0, -2.5, -3.0$, and $-3.5$ \citep{demarque2004}. As the stars are selected to be bright
and cool they mostly occupy the upper portion of the giant branch in this diagram. Four stars that are somewhat hotter than the rest of the sample were also observed, mainly because they were bright and/or due to the lack of other available targets in the appropriate RA range. Some of our stars also turned out to be more metal-rich than our initial selection criteria. This is mainly because not quite all of the targets initially chosen passed through the Phase-I medium-resolution vetting process described in section \ref{sec:strategy}. A separate paper will address differences in stellar parameters determined from the medium- and high-resolution spectra, respectively. 

\begin{deluxetable*}{llrrrrr}
\tablecaption{Equivalent Widths of Fe~I and Fe~II lines in our Sample \label{tab:ew}}
\tablewidth{0pt}
\tablehead{\colhead{Stellar ID} & \colhead{Species} & \colhead{$\lambda$} & \colhead{$\chi$} & \colhead{$\log$ $gf$} & \colhead{EW} & \colhead{$\log \epsilon$}\\ 
\colhead{} & \colhead{} & \colhead{(\AA)} & \colhead{(eV)} & \colhead{} & \colhead{(m\AA)} & \colhead{}}
\startdata
J00002259$-$1302275  &  \ion{Fe}{1}  &  4143.87  &  1.56  &  $-$0.51  &  133.30  &  5.04 \\
J00002259$-$1302275  &  \ion{Fe}{1}  &  4191.43  &  2.47  &  $-$0.67  &  94.10  &  5.28 \\
J00002259$-$1302275  &  \ion{Fe}{1}  &  4415.12  &  1.61  &  $-$0.62  &  117.10  &  4.78 \\
J00002259$-$1302275  &  \ion{Fe}{1}  &  4447.72  &  2.22  &  $-$1.34  &  94.40  &  5.59 \\
J00002259$-$1302275  &  \ion{Fe}{1}  &  4461.65  &  0.09  &  $-$3.19  &  122.20  &  5.73 \\
\multicolumn{1}{c}{\vdots} & \multicolumn{1}{l}{\phantom{0}\vdots} & \multicolumn{1}{c}{\vdots} & \multicolumn{1}{c}{\vdots} & \multicolumn{1}{c}{\vdots} & \multicolumn{1}{c}{\vdots} & \multicolumn{1}{c}{\vdots} \\
J00002259$-$1302275  &  \ion{Fe}{1}  &  6593.87  &  2.43  &  $-$2.42  &  30.50  &  5.44 \\
J00002259$-$1302275  &  \ion{Fe}{1}  &  6750.15  &  2.42  &  $-$2.62  &  43.30  &  5.87 \\
J00002259$-$1302275  &  \ion{Fe}{2}  &  4522.63  &  2.84  &  $-$1.99  &  83.00  &  5.35 \\
J00002259$-$1302275  &  \ion{Fe}{2}  &  4620.52  &  2.83  &  $-$3.19  &  74.20  &  6.33 \\
J00002259$-$1302275  &  \ion{Fe}{2}  &  4923.93  &  2.89  &  $-$1.21  &  87.50  &  4.68 \\
J00002259$-$1302275  &  \ion{Fe}{2}  &  5234.63  &  3.22  &  $-$2.18  &  37.00  &  4.97 \\
J00002259$-$1302275  &  \ion{Fe}{2}  &  5316.62  &  3.15  &  $-$1.78  &  94.10  &  5.65 \\
J00002259$-$1302275  &  \ion{Fe}{2}  &  6432.68  &  2.89  &  $-$3.71  &  34.70  &  6.04 \\
J00021222$-$2241388  &  \ion{Fe}{1}  &  4021.87  &  2.67  &  $-$0.73  &  84.24  &  5.43 \\
J00021222$-$2241388  &  \ion{Fe}{1}  &  4127.78  &  3.28  &  $-$1.61  &  18.42  &  5.55 \\
J00021222$-$2241388  &  \ion{Fe}{1}  &  4132.90  &  2.84  &  $-$0.92  &  58.74  &  5.22 \\
J00021222$-$2241388  &  \ion{Fe}{1}  &  4154.50  &  2.83  &  $-$0.69  &  76.67  &  5.41 \\
\multicolumn{1}{c}{\vdots} & \multicolumn{1}{l}{\phantom{0}\vdots} & \multicolumn{1}{c}{\vdots} & \multicolumn{1}{c}{\vdots} & \multicolumn{1}{c}{\vdots} & \multicolumn{1}{c}{\vdots} & \multicolumn{1}{c}{\vdots} \\
\enddata
\tablecomments{
The complete version of Table~\ref{tab:ew} is available online only.
A short version is shown here to illustrate its form and content.
}
\end{deluxetable*}
\startlongtable
\begin{deluxetable*}{lrrrrrrr}
\tablecaption{Stellar Parameters \label{tab:param}}
\tablewidth{0pt}
\tablehead{
\colhead{Stellar ID} & \colhead{$T_{\rm eff}$} & \colhead{$\log g$} & \colhead{$\mathrm{[Fe/H]}$}& \colhead{$\sigma_{\mathrm{[Fe/H]}}$} &\colhead{$N$(\ion{Fe}{1})} &\colhead{$N$(\ion{Fe}{2})} &\colhead{$\xi$}\\
\colhead{} & \colhead{(K)} & \colhead{} & \colhead{}& \colhead{(dex)}
&\colhead{} &\colhead{} &\colhead{(km~s$^{-1}$)}}
\startdata 
J00002259-1302275	&4576	&2.12	&$-$2.90	&0.21 & 58	&6	&2.12\\
J00021222-2241388	&5035	&2.09	&$-$2.19	&0.19 & 107	&13	&1.85\\
J00021668-2453494	&5020	&1.73	&$-$1.81	&0.19 &	85	&19 &1.85\\
J00133067-1259594	&4405	&0.50	&$-$2.82	&0.17 & 71	&14	&2.60\\
J00233067-1631428	&5382	&3.04	&$-$2.45	&0.14 & 87	&10	&1.74\\
J00400685-4325183   &4630	&1.16	&$-$2.55	&0.21 & 64	&9	&2.75\\
J00405260-5122491	&5689	&4.51	&$-$2.11	&0.09 & 97	&11	&1.61\\
J00453930-7457294	&4947	&2.01	&$-$2.00	&0.16 & 115	&22	&1.90\\
J01202234-5425582	&5125	&2.53	&$-$2.11	&0.16 & 104	&22	&1.75\\
J01293113-1600454	&4876	&1.80	&$-$2.81	&0.12 & 74	&14	&2.13\\
J01334657-2727374	&5227	&2.79	&$-$1.60	&0.15 & 56	&7	&1.95\\
J01425422-5032488	&5132	&2.65	&$-$2.09	&0.12 & 122	&20	&1.55\\
J01430726-6445174	&4590	&0.92	&$-$3.00	&0.12 & 97	&13	&2.12\\
J01451951-2800583	&4495	&0.67	&$-$2.80	&0.15 & 69	&20	&2.10\\
J01490794-4911429	&4682	&0.93	&$-$2.94	&0.11 & 78	&21	&1.95\\
J01530024-3417360	&5323	&3.26	&$-$1.50	&0.13 & 132	&21	&1.40\\
J02040793-3127556	&4550	&0.92	&$-$3.08	&0.17 & 105	&20	&2.35\\
J02165716-7547064	&4543	&0.71	&$-$2.50	&0.14 & 90	&16	&2.44\\
J02355867-6745520	&4653	&0.91	&$-$1.55	&0.20 & 58	&14	&2.04\\
J02401075-1416290	&6070	&4.45	&$-$0.79	&0.19 & 47	&16	&0.80\\
J02412152-1825376	&4405	&0.80	&$-$2.17	&0.14 & 71	&19	&1.90\\
J02441479-5158241	&4869	&1.87	&$-$2.93	&0.13 & 86	&16	&1.69\\
J02462013-1518419	&4879	&1.80	&$-$2.71	&0.17 & 69	&13	&2.35\\
J02500719-5145148	&4707	&1.40	&$-$2.20	&0.20 & 115	&23	&2.30\\
J03193531-3250433	&6311	&3.05	&$-$2.93	&0.17 & 30	&7	&0.81\\
J03563703-5838281	&4450	&0.70	&$-$2.87	&0.16 & 68	&12	&2.60\\
J04121388-1205050	&4658	&0.95	&$-$2.73	&0.16 & 29	&11	&2.58\\
J13164824-2743351	&4990	&2.24	&$-$1.61	&0.17 & 127	&20	&1.50\\
J14033542-3335257	&4889	&1.75	&$-$2.71	&0.14 & 132	&15	&1.77\\
J14164084-2422000	&4540	&0.57	&$-$2.73	&0.13 & 77	&23	&1.95\\
J14232679-2834200	&5200	&2.66	&$-$1.90	&0.19 & 100	&12	&1.65\\
J14301385-2317388	&4490	&1.25	&$-$1.40	&0.16 & 92	&18	&2.40\\
J14325334-4125494	&5020	&2.39	&$-$2.79	&0.16 & 108	&12	&1.80\\
J14381119-2120085	&4608	&1.20	&$-$2.21	&0.20 & 109	&18	&2.20\\
J15260106-0911388	&4499	&0.76	&$-$2.83	&0.16 & 81	&14	&2.41\\
J15271353-2336177	&5882	&4.01	&$-$2.15	&0.12 & 130	&17	&1.35\\
J15575183-0839549	&4730	&1.88	&$-$1.49	&0.16 & 109	&18	&1.87\\
J15582962-1224344	&5125	&1.52	&$-$2.54	&0.16 & 105	&22	&2.35\\
J16024498-1521016	&5240	&3.16	&$-$1.80	&0.17 & 94	&22	&1.65\\
J16080681-3215592	&4680	&1.27	&$-$2.07	&0.16 & 119	&23	&1.98\\
J16095117-0941174	&4556	&0.68	&$-$2.82	&0.12 & 86	&21	&2.18\\
J16103106+1003055	&5370	&2.60	&$-$2.43	&0.19 & 64	&6	&2.05\\
J16184302-0630558	&5407	&1.67	&$-$2.60	&0.10 & 66	&16	&2.19\\
J17093199-6027271	&4547	&0.77	&$-$2.48	&0.20 & 70	&13	&2.71\\
J17094926-6239285	&4585	&0.90	&$-$2.51	&0.15 & 73	&14	&2.00\\
J17124284-5211479	&4750	&1.15	&$-$2.78	&0.19 & 74	&15	&2.00\\
J17225742-7123000	&5080	&2.67	&$-$2.42	&0.18 & 58	&12	&0.70\\
J17255680-6603395	&4686	&0.69	&$-$2.59	&0.11 & 80	&14	&2.38\\
J17273886-5133298	&4992	&1.38	&$-$1.78	&0.14 & 95	&22	&1.91\\
J17285930-7427532	&4900	&1.61	&$-$2.04	&0.17 & 74	&13	&1.80\\
J17334285-5121550	&4899	&1.18	&$-$1.91	&0.14 & 104	&22	&2.08\\
J17400682-6102129	&4880	&1.78	&$-$2.24	&0.16 & 107	&18	&2.08\\
J17435113-5359333	&5080	&1.65	&$-$2.24	&0.16 & 124	&20	&2.10\\
J18024226-4404426	&4701	&1.60	&$-$1.55	&0.18 & 88	&14	&2.17\\
J18121045-4934495	&4436	&0.76	&$-$3.02	&0.14 & 66	&13	&2.95\\
J18285086-3434203	&4630	&1.12	&$-$2.46	&0.15 & 118	&17	&2.15\\
J18294122-4504000	&4700	&1.25	&$-$2.48	&0.17 & 80	&11	&2.10\\
J18362318-6428124	&4896	&1.80	&$-$2.57	&0.10 & 97	&19	&1.95\\
J18363613-7136597	&4650	&0.65	&$-$2.52	&0.16 & 77	&20	&2.45\\
J18562774-7251331	&4709	&1.08	&$-$2.26	&0.15 & 97	&8	&2.15\\
J19014952-4844359	&4820	&1.84	&$-$1.87	&0.16 & 94	&9	&1.75\\
J19161821-5544454	&4450	&0.65	&$-$2.35	&0.18 & 66	&18	&2.50\\
J19172402-4211240	&4775	&1.06	&$-$2.61	&0.15 & 119	&18	&2.40\\
J19173310-6628545	&4498	&1.05	&$-$2.76	&0.15 & 64	&12	&2.72\\
J19202070-6627202	&4690	&1.64	&$-$2.10	&0.18 & 111	&17	&1.70\\
J19215077-4452545	&4450	&0.10	&$-$2.56	&0.17 & 74	&11	&2.30\\
J19232518-5833410	&5035	&2.12	&$-$2.08	&0.18 & 82	&8	&2.05\\
J19324858-5908019	&4540	&1.92	&$-$1.93	&0.18 & 46	&11	&1.70\\
J19345497-5751400	&4580	&1.74	&$-$2.46	&0.14 & 28	&7	&2.50\\
J19494025-5424113	&4486	&0.91	&$-$2.81	&0.17 & 72	&13	&2.24\\
J19534978-5940001	&4614	&0.70	&$-$2.79	&0.16 & 66	&13	&2.41\\
J19594558-2549075	&4540	&0.79	&$-$2.44	&0.15 & 95	&21	&2.64\\
J20093393-3410273	&4690	&1.64	&$-$2.10	&0.18 & 111	&17	&1.70\\
J20144608-5635300	&4765	&1.19	&$-$1.92	&0.18 & 41	&18	&1.80\\
J20165357-0503592   &4585   &0.90   &$-$2.89    &0.18 & 88	&10   &2.90\\
J20303339-2519500	&4315	&0.60	&$-$2.18	&0.18 & 106	&24	&2.30\\
J20313531-3127319	&4894	&1.39	&$-$2.43	&0.14 & 82	&13	&1.83\\
J20445584-2250597	&4366	&1.00	&$-$2.30	&0.18 & 111	&15	&2.16\\
J20492765-5124440	&4250	&1.09	&$-$2.47	&0.19 & 80	&11	&2.60\\
J20514971-6158008	&5285	&3.13	&$-$1.87	&0.14 & 114	&19	&1.68\\
J20542346-0033097	&4351	&0.68	&$-$2.78	&0.20 & 82	&11	&2.64\\
J20560913-1331176	&4780	&1.60	&$-$2.30	&0.13 & 119	&24	&1.80\\
J21023752-5316132	&5045	&2.40	&$-$2.44	&0.19 & 72	&13	&1.95\\
J21063474-4957500	&5238	&2.82	&$-$2.91	&0.13 & 102	&15	&1.93\\
J21064294-6828266	&5186	&2.70	&$-$2.76	&0.19 & 43	&10	&2.90\\
J21091825-1310062	&4855	&1.42	&$-$2.40	&0.17 & 106	&24	&2.10\\
J21095804-0945400	&4492	&0.98	&$-$2.73	&0.16 & 57	&10	&2.11\\
J21141350-5726363	&4690	&1.21	&$-$2.69	&0.15 & 140	&10	&2.05\\
J21152551-1503309	&4722	&1.47	&$-$2.06	&0.17 & 68	&18	&2.00\\
J21162185-0213420	&4390	&0.73	&$-$2.23	&0.16 & 62	&11	&1.90\\
J21224590-4641030	&4652	&1.42	&$-$2.96	&0.14 & 67	&14	&2.78\\
J21262525-2144243	&4450	&1.09	&$-$2.81	&0.17 & 121	&12	&2.45\\
J21291592-4236219	&4650	&1.15	&$-$2.48	&0.19 & 96	&18	&2.10\\
J21370807-0927347	&4470	&0.99	&$-$2.46	&0.18 & 84	&14	&2.05\\
J21513595-0543398	&4530	&0.66	&$-$2.46	&0.15 & 99	&27	&2.21\\
J22021636-0536483	&4668	&0.93	&$-$2.75	&0.18 & 67	&14	&2.57\\
J22163596+0246171	&4900	&1.92	&$-$2.37	&0.18 & 131	&10	&1.85\\
J22444701-5617540	&4495	&1.41	&$-$1.86	&0.21 & 43	&11	&1.75\\
J22492756-2238289	&4930	&1.81	&$-$1.82	&0.19 & 78	&10	&2.35\\
J22531984-2248555	&4720	&1.59	&$-$1.87	&0.18 & 91	&18	&1.65\\
J22595884-1554182	&4600	&0.92	&$-$2.64	&0.16 & 91	&17	&2.55\\
J23022289-6833233	&5063	&2.29	&$-$2.64	&0.15 & 110	&14	&2.21\\
J23044914-4243477	&4733	&2.12	&$-$2.43	&0.28 & 70	&13	&1.25\\
J23130003-4507066	&4612	&0.95	&$-$2.64	&0.18 & 87	&18	&2.21\\
J23265258-0159248	&4360	&0.45	&$-$3.13	&0.16 & 82	&21	&2.40\\
J23310716-0223301	&5050	&1.70	&$-$2.28	&0.15 & 30	&7	&2.05\\
J23362202-5607498	&4630	&1.28	&$-$2.06	&0.20 & 38	&14	&2.15\\
\enddata
\end{deluxetable*}

\subsection{Abundances}

As described in section \ref{sec:strategy}, we derive abundances for C, Sr, Ba, and Eu for the stars in our sample, which allows us to classify the stars as either $r$-I, $r$-II, limited-$r$, CEMP-$s$, CEMP-$r$, or non-$r$-process-enhanced. As the stars were preferentially selected not to exhibit carbon enhancement, we did not expect to find a large number of CEMP-$s$ or CEMP-$r$ stars, but nevertheless included carbon measurements in our analysis. A full abundance analysis of the stars will follow in future papers. All abundances are derived
from spectral syntheses using MOOG2017 and $\alpha$-enhanced
($\mathrm{[\alpha/Fe]} = +0.4$) 1D LTE ATLAS9 model atmospheres
\citep{castelli2003}. We used Solar photosphere abundances from
\cite{asplund2009} and line lists generated using the linemake
package\footnote{https://github.com/vmplacco/linemake} (C.\ Sneden, private comm.), including molecular lines from CH, C$_2$, and CN
\citep{brooke2013,masseron2014,ram2014,sneden2014} and isotopic shift and hyperfine structure information for Ba and Eu
\citep{lawler2001,gallagher2010}. 

\begin{figure*}
\center
\includegraphics[width=0.8\textwidth]{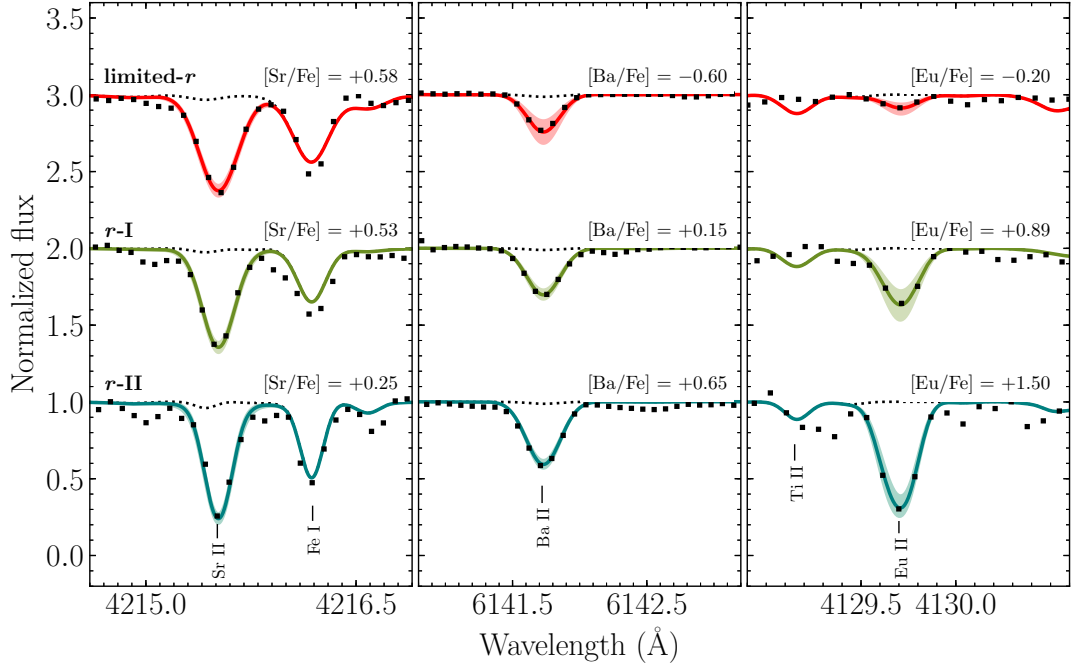}
\caption{Syntheses of the \ion{Sr}{2}~4215\,{\AA} (left), \ion{Ba}{2}~6141\,{\AA} (middle), and \ion{Eu}{2}~4129\,{\AA} (right) features for a limited-$r$ star (J18121045$-$4934495; red, top), an $r$-I star (J15582962$-$1224344; green, middle), and an $r$-II star (J02462013$-$1518419; blue, bottom).
Following the representative errors given in Table \ref{tab:error}, the \ion{Sr}{2} and \ion{Ba}{2} syntheses are shown with a $\pm$0.30~dex uncertainty region and \ion{Eu}{2} with a $\pm$0.20~dex uncertainty (filled).
Dashed lines show syntheses lacking the respective element.
\label{fig:spec}}  
\end{figure*}

Carbon abundances are derived by fitting the CH $G$-band at 4313\,{\AA}
and/or the C$_{2}$ Swan band at 5161\,{\AA}. Three Sr lines are available in our wavelength range at 4077\,{\AA}, 4161\,{\AA}, and 4215\,{\AA}. The 4077\,{\AA} line is very strong and is often saturated in our spectra; on the other hand, the 4161\,{\AA} is weak and severely blended. Hence, our Sr abundances are primarily derived from the line at 4215\,{\AA}. This line can be blended with CN features in cool metal-poor stars. However, as our stars are chosen not to be enhanced in C, this was only a complication for a few spectra. Barium abundances are derived from three of the five Ba lines in our wavelength range at 5853\,{\AA}, 6141\,{\AA}, and 6496\,{\AA}. There are also Ba lines at 4554\,{\AA} and 4934\,{\AA}, but they were often saturated in these cool stars. Furthermore, the 4934\,{\AA} line is also blended with an Fe~I feature for which the  $\log gf$ value is poorly constrained; these two lines were disregarded for the final Ba abundance. For Eu, eight absorption lines are present in our spectral range, at 3819\,{\AA}, 3907\,{\AA}, 4129\,{\AA}, 4205\,{\AA}, 4435\,{\AA}, 4522\,{\AA}, 6437\,{\AA}, and 6645\,{\AA}. The 3819\,{\AA} and 3907\,{\AA} lines are in a region of our spectra with poor S/N, and the 4205\,{\AA}, 4435\,{\AA}, and 4522\,{\AA} features are heavily blended. Hence, our Eu abundances are mainly derived from the 4129\,{\AA} feature, and those at 6437\,{\AA} and 6645\,{\AA}, when detectable. Examples of the synthesis of Sr, Ba, and Eu lines used for the classification of a limited-$r$ star, an $r$-I star, and an $r$-II star are shown in Figure~\ref{fig:spec}.

\section{Results}
\label{results}

\startlongtable
\begin{deluxetable*}{lrrrrrrrl}
\tablecaption{Abundances \label{tab:abun}}
\tablewidth{0pt}
\tablehead{
\colhead{Stellar ID} & \colhead{$\mathrm{[Fe/H]}$} & \colhead{$\mathrm{[C/Fe]}$} & \colhead{$\mathrm{[Sr/Fe]}$} & \colhead{$\mathrm{[Ba/Fe]}$} & \colhead{$\mathrm{[Eu/Fe]}$}
& \colhead{$\mathrm{[Ba/Eu]}$} & \colhead{$\mathrm{[Sr/Ba]}$} & \colhead{Sub-class} }
\startdata 
J00002259$-$1302275	&$-$2.90	&$-$0.65	&$-$1.20	&$-$0.38	&$+$0.58	&$-$0.96	&$-$0.82 &$r$-I\\
J00021222$-$2241388	&$-$2.19	&$-$0.19	&$-$0.13	&$-$0.41	&$+$0.22	&$-$0.63	&$+$0.28 &\\
J00021668$-$2453494	&$-$1.81	&$-$0.88	&$+$0.59	&$+$0.10	&$+$0.52	&$-$0.42	&$+$0.49 &$r$-I\\
J00133067$-$1259594	&$-$2.82	&$-$0.58	&$-$0.25	&$-$0.55	&$-$0.06	&$-$0.49	&$+$0.30 &\\
J00233067$-$1631428	&$-$2.45	&$+$0.37	&$-$0.75	&$-$0.55	&$<+$0.27	&$>-$0.82	&$-$0.20 & Unknown\\
J00400685$-$4325183 &$-$2.55	&$-$0.85	&$-$1.52	&$-$1.56	&$+$0.55	&$-$2.11	&$+$0.04 &$r$-I\\
J00405260$-$5122491	&$-$2.11	&$-$0.04	&$+$0.09	&$-$0.04	&$+$0.86	&$-$0.90	&$+$0.13 &$r$-I\\
J00453930$-$7457294	&$-$2.00	&$+$0.93	&$+$0.83	&$+$0.37	&$+$0.55	&$-$0.18	&$+$0.46 &CEMP-$r$\\
J01202234$-$5425582	&$-$2.11	&$-$0.09	&$+$0.50	&$+$0.16	&$+$0.30	&$-$0.14	&$+$0.34 &$r$-I\\
J01293113$-$1600454	&$-$2.81	&$+$0.35	&$+$0.88	&$+$0.95	&$+$1.76	&$-$0.81	&$-$0.07 &$r$-II\\
J01334657$-$2727374	&$-$1.60	&$+$1.64	&$+$1.40	&$+$1.61	&$+$0.88	&$+$0.73	&$-$0.21 &CEMP-$s$\\
J01425422$-$5032488	&$-$2.09	&$+$0.07	&$+$0.13	&$-$0.13	&$+$0.38	&$-$0.51	&$+$0.26 &$r$-I\\
J01430726$-$6445174	&$-$3.00	&$-$0.14	&$-$1.00	&$-$0.51	&$-$0.14	&$-$0.37	&$-$0.49 &\\
J01451951$-$2800583	&$-$2.80	&$-$0.69	&$+$0.30	&$-$0.98	&$<-$0.38	&$>-$0.60	&$+$1.28 &limited-$r$\\	
J01490794$-$4911429	&$-$2.94	&$+$0.05	&$-$0.45	&$-$0.59	&$+$0.09	&$-$0.68	&$+$0.14 &\\
J01530024$-$3417360	&$-$1.50	&$+$0.01	&$+$0.40	&$+$0.09	&$+$0.71	&$-$0.62	&$+$0.31 &$r$-I\\
J02040793$-$3127556	&$-$3.08	&$-$0.76	&$+$0.07	&$-$0.77	&$<+$0.00	&$>-$0.77	&$+$0.84 &limited-$r$\\
J02165716$-$7547064	&$-$2.50	&$-$0.33	&$+$0.25	&$+$0.25	&$+$1.12	&$-$0.87	&$+$0.00 &$r$-II\\
J02355867$-$6745520	&$-$1.55	&$+$0.97	&$+$0.10	&$+$0.49	&$+$0.50	&$-$0.01	&$-$0.39 &CEMP-$r$\\
J02401075$-$1416290	&$-$0.79	&$+$1.20	&$>+$1.00	&$+$1.69	&$+$0.84	&$+$0.85	&$>-$0.69 &CEMP-$s$\\
J02412152$-$1825376	&$-$2.17	&$-$0.96	&$+$0.35	&$+$0.09	&$+$0.20	&$-$0.11	&$+$0.26 &\\
J02441479$-$5158241	&$-$2.93	&$-$0.05	&$-$0.55	&$-$0.80	&$+$0.20	&$-$1.00	&$+$0.25 &\\
J02462013$-$1518419	&$-$2.71	&$+$0.04	&$+$0.58	&$+$0.60	&$+$1.45	&$-$0.85	&$-$0.02 &$r$-II\\
J02500719$-$5145148	&$-$2.20	&$+$1.04	&$-$0.05	&$+$0.48	&$+$0.62	&$-$0.14	&$-$0.53 &CEMP-$r$\\
J03193531$-$3250433	&$-$2.93	&$<+$1.60	&$-$0.28	&$<-$0.15	&$<+$1.14	&\nodata    &$>-$0.13 & Unknown\\	
J03563703$-$5838281	&$-$2.87	&$-$0.78	&$+$0.55	&$-$0.50	&$+$0.11	&$-$0.61	&$+$1.05 &limited-$r$\\
J04121388$-$1205050	&$-$2.73	&$-$0.12	&$+$0.42	&$-$0.43	&$+$0.52	&$-$0.95	&$+$0.85 &$r$-I\\
J13164824$-$2743351	&$-$1.61	&$-$0.05	&$+$0.26	&$+$0.01	&$+$0.54	&$-$0.53	&$+$0.25 &$r$-I\\
J14033542$-$3335257	&$-$2.71	&$+$0.08	&$+$0.10	&$-$0.28	&$+$0.29	&$-$0.57	&$+$0.38 &\\
J14164084$-$2422000	&$-$2.73	&$-$0.86	&$+$0.48	&$-$0.44	&$+$0.02	&$-$0.46	&$+$0.92 &limited-$r$\\
J14232679$-$2834200	&$-$1.90	&$+$0.43	&$+$0.44	&$-$0.07	&$+$0.61	&$-$0.68	&$+$0.51 &$r$-I\\
J14301385$-$2317388	&$-$1.40	&$-$0.70	&$+$0.40	&$+$0.34	&$+$0.62	&$-$0.28	&$+$0.06 &$r$-I\\
J14325334$-$4125494	&$-$2.79	&$-$0.14	&$+$0.32	&$+$0.71	&$+$1.61	&$-$0.90	&$-$0.39 &$r$-II\\
J14381119$-$2120085	&$-$2.21	&$+$0.70	&$+$0.42	&$+$0.97	&$+$0.62	&$+$0.35	&$-$0.55 &\\
J15260106$-$0911388	&$-$2.83	&$-$0.82	&$+$0.90	&$+$0.69	&$+$1.70	&$-$1.01	&$+$0.21 &$r$-II\\
J15271353$-$2336177	&$-$2.15	&$+$0.33	&$+$0.35	&$-$0.03	&$+$0.70	&$-$0.73	&$+$0.38 &$r$-I\\
J15575183$-$0839549	&$-$1.49	&$+$0.60	&$+$1.20	&$+$1.68	&$+$0.70	&$+$0.98	&$-$0.48 &\\
J15582962$-$1224344	&$-$2.54	&$-$0.14	&$+$0.53	&$+$0.04	&$+$0.89	&$-$0.85	&$+$0.49 &$r$-I\\
J16024498$-$1521016	&$-$1.80	&$-$0.05	&$+$0.35	&$+$0.08	&$+$0.55	&$-$0.47	&$+$0.27 &$r$-I\\
J16080681$-$3215592	&$-$2.07	&$-$0.32	&$+$0.07	&$+$0.08	&$+$0.25	&$-$0.17	&$-$0.01 &\\
J16095117$-$0941174	&$-$2.82	&$-$0.65	&$-$0.10	&$-$0.29	&$+$0.17	&$-$0.46	&$+$0.19 &\\
J16103106$+$1003055	&$-$2.43	&$+$0.53	&$-$0.19	&$-$1.10	&$<+$0.70	&$>-$1.80   &$+$0.91 &Unknown\\
J16184302$-$0630558	&$-$2.60	&$+$0.40	&$-$0.20	&$<-$0.60	&$<+$0.20	&\nodata	&$>+$0.40 &Unknown\\	
J17093199$-$6027271	&$-$2.47	&$-$0.45	&$+$0.28	&$-$0.15	&$+$0.60	&$-$0.75	&$+$0.43 &$r$-I\\
J17094926$-$6239285	&$-$2.51	&$-$0.83	&$-$0.05	&$-$0.45	&$+$0.02	&$-$0.47	&$+$0.40 &\\
J17124284$-$5211479	&$-$2.78	&$-$0.42	&$-$0.03	&$-$0.58	&$+$0.48	&$-$1.06	&$+$0.55 &$r$-I\\
J17225742$-$7123000	&$-$2.42	&$-$0.33	&$+$0.31	&$+$0.70	&$+$1.07	&$-$0.37	&$-$0.39 &$r$-II\\
J17255680$-$6603395	&$-$2.59	&$-$0.42	&$+$0.04	&$-$0.35	&$+$0.18	&$-$0.53	&$+$0.39 &\\
J17273886$-$5133298	&$-$1.78	&$-$0.45	&$+$0.12	&$+$0.48	&$+$0.48	&$+$0.00	&$-$0.36 &\\
J17285930$-$7427532	&$-$2.04	&$-$0.42	&$+$0.74	&$-$0.53	&$+$0.23	&$-$0.76	&$+$1.27 &limited-$r$\\
J17334285$-$5121550	&$-$1.91	&$-$0.32	&$+$0.33	&$+$0.19	&$+$0.20	&$-$0.01	&$+$0.14 &\\
J17400682$-$6102129	&$-$2.24	&$-$0.26	&$+$0.16	&$-$0.31	&$+$0.31	&$-$0.62	&$+$0.47 &$r$-I\\
J17435113$-$5359333	&$-$2.24	&$-$0.35	&$+$0.00	&$-$0.07	&$+$0.73	&$-$0.80	&$+$0.07 &$r$-I\\
J18024226$-$4404426	&$-$1.55	&$+$0.35	&$+$0.68	&$+$0.95	&$+$1.05	&$-$0.10	&$-$0.27 &$r$-II\\
J18121045$-$4934495	&$-$3.02	&$-$0.33	&$+$0.25	&$-$0.50	&$-$0.07	&$-$0.43	&$+$0.75 &limited-$r$\\
J18285086$-$3434203	&$-$2.46	&$-$0.58	&$+$0.22	&$-$0.29	&$+$0.53	&$-$0.82	&$+$0.51 &$r$-I\\
J18294122$-$4504000	&$-$2.48	&$-$0.46	&$+$0.47	&$-$0.06	&$+$0.70	&$-$0.76	&$+$0.53 &$r$-I\\
J18362318$-$6428124	&$-$2.57	&$+$0.10	&$-$0.63	&$+$0.18	&$+$0.57	&$-$0.39	&$-$0.81 &$r$-I\\
J18363613$-$7136597	&$-$2.52	&$-$0.75	&$+$0.01	&$-$0.37	&$+$0.56	&$-$0.93	&$+$0.38 &$r$-I\\
J18562774$-$7251331	&$-$2.26	&$-$0.48	&$+$0.00	&$-$0.01	&$+$0.32	&$-$0.33	&$+$0.01 &$r$-I\\
J19014952$-$4844359	&$-$1.87	&$+$0.02	&$+$0.37	&$-$0.24	&$+$0.93	&$-$1.17	&$+$0.61 &$r$-I\\
J19161821$-$5544454	&$-$2.35	&$-$0.80	&$+$0.48	&$-$0.12	&$+$1.08	&$-$1.20	&$+$0.60 &$r$-II\\
J19172402$-$4211240	&$-$2.44	&$-$0.58	&$-$0.13	&$-$0.33	&$+$0.11	&$-$0.44	&$+$0.20 &\\
J19173310$-$6628545	&$-$2.76	&$-$0.40	&$-$0.35	&$-$0.60	&$+$0.00	&$-$0.60	&$+$0.25 &\\
J19202070$-$6627202	&$-$2.10	&$-$0.28	&$+$0.73	&$-$0.11	&$+$0.29	&$-$0.40	&$+$0.84 &limited-$r$\\
J19215077$-$4452545	&$-$2.56	&$-$0.77	&$-$0.05	&$-$0.31	&$+$0.74	&$-$1.05	&$+$0.26 &$r$-I\\
J19232518$-$5833410	&$-$2.08	&$+$0.27	&$+$0.56	&$+$0.11	&$+$0.76	&$-$0.65	&$+$0.45 &$r$-I\\
J19324858$-$5908019	&$-$1.93	&$-$0.51	&$+$0.90	&$+$0.65	&$+$0.90	&$-$0.25	&$+$0.25 &$r$-I\\
J19345497$-$5751400	&$-$2.46	&$-$0.70	&$+$0.25	&$-$0.71	&$+$0.17	&$-$0.88	&$+$0.96 &limited-$r$\\
J19494025$-$5424113	&$-$2.81	&$-$0.55	&$-$0.05	&$-$1.01	&$-$0.39	&$-$0.62	&$+$0.96 &limited-$r$\\
J19534978$-$5940001	&$-$2.79	&$-$0.44	&$+$0.25	&$-$0.52	&$+$0.04	&$-$0.56	&$+$0.77 &limited-$r$\\
J19594558$-$2549075	&$-$2.44	&$-$0.70	&$+$0.10	&$-$0.50	&$+$0.04	&$-$0.54	&$+$0.60 &limited-$r$\\
J20093393$-$3410273	&$-$1.99	&$-$0.86	&$+$0.59	&$+$0.23	&$+$1.32	&$-$1.09	&$+$0.36 &$r$-II\\
J20144608$-$5635300	&$-$1.92	&$+$1.50	&$+$0.98	&$+$1.08	&$+$0.92	&$+$0.16	&$-$0.10 &CEMP-$s$\\
J20165357$-$0503592 &$-$2.89	&$-$0.12	&$-$2.43	&$-$1.10    &$<+$0.00	&$>-$1.10   &$-$1.33 & \\
J20303339$-$2519500	&$-$2.21	&$-$0.65	&$+$0.41	&$-$0.29	&$+$0.41	&$-$0.70	&$+$0.70 &$r$-I\\
J20313531$-$3127319	&$-$2.43	&$+$0.10	&$+$0.20	&$-$0.79	&$<-$0.30	&$>-$0.49	&$+$0.99 &limited-$r$\\
J20445584$-$2250597	&$-$2.30	&$-$0.57	&$+$0.40	&$-$0.05	&$-$0.09	&$+$0.04	&$+$0.45 &\\
J20492765$-$5124440	&$-$2.47	&$-$0.77	&$+$0.13	&$-$0.25	&$+$0.58	&$-$0.83	&$+$0.38 &$r$-I\\
J20514971$-$6158008	&$-$1.87	&$+$0.10	&$+$0.28	&$-$0.18	&$+$0.43	&$-$0.61	&$+$0.46 &$r$-I\\
J20542346$-$0033097	&$-$2.78	&$-$0.36	&$+$0.86	&$-$0.03	&$<+$0.00	&$>-$0.03	&$+$0.89 &limited-$r$\\
J20560913$-$1331176	&$-$2.30	&$+$0.00	&$+$0.13	&$-$0.56	&$+$0.00	&$-$0.56	&$+$0.69 &limited-$r$\\
J21023752$-$5316132	&$-$2.44	&$-$0.38	&$-$0.42	&$-$1.63	&$<+$0.20	&$>-$1.43	&$+$1.21 &limited-$r$\\
J21063474$-$4957500	&$-$2.91	&$+$0.52	&$-$0.48	&$-$0.20	&$+$0.54	&$-$0.74	&$-$0.28 &$r$-I\\
J21064294$-$6828266	&$-$2.76	&$+$0.53	&$+$0.90	&$+$0.52	&$+$1.32	&$-$0.80	&$+$0.38 &$r$-II\\
J21091825$-$1310062	&$-$2.40	&$-$0.28	&$+$0.14	&$+$0.12	&$+$1.25	&$-$1.13	&$+$0.02 &$r$-II\\
J21095804$-$0945400	&$-$2.73	&$-$0.45	&$+$0.60	&$-$0.12	&$+$0.77	&$-$0.89	&$+$0.72 &$r$-I\\
J21141350$-$5726363	&$-$2.69	&$-$0.10	&$-$0.32	&$-$0.94	&$<-$0.40	&$>-$0.54	&$+$0.62 &limited-$r$\\
J21152551$-$1503309	&$-$2.06	&$-$0.07	&$+$1.93	&$+$1.41	&$+$0.73	&$+$0.68	&$+$0.52 &\\
J21162185$-$0213420	&$-$2.23	&$-$0.98	&$+$0.57	&$+$0.06	&$+$0.49	&$-$0.43	&$+$0.51 &$r$-I\\
J21224590$-$4641030	&$-$2.96	&$-$0.18	&$+$0.25	&$+$0.01	&$+$0.90	&$-$0.89	&$+$0.24 &$r$-I\\
J21262525$-$2144243	&$-$2.81	&$-$0.88	&$+$0.37	&$-$0.02	&$+$0.58	&$-$0.60	&$+$0.39 &$r$-I\\
J21291592$-$4236219	&$-$2.48	&$-$0.07	&$-$0.25	&$-$0.54	&$<-$0.20	&$>-$0.34	&$+$0.29 &Unknown\\
J21370807$-$0927347	&$-$2.46	&$-$0.50	&$+$0.63	&$-$0.61	&$+$0.11	&$-$0.72	&$+$1.24 &limited-$r$\\
J21513595$-$0543398	&$-$2.46	&$-$1.20	&$-$0.30	&$-$0.66	&$+$0.02	&$-$0.68	&$+$0.36 &\\
J22021636$-$0536483	&$-$2.75	&$+$0.05	&$+$0.15	&$-$0.75	&$<-$0.12	&$>-$0.63	&$+$0.90 &limited-$r$\\	
J22163596$+$0246171	&$-$2.37	&$-$0.10	&$+$0.17	&$-$0.27	&$+$0.45	&$-$0.72	&$+$0.44 &$r$-I\\
J22444701$-$5617540	&$-$1.86	&$-$0.43	&$+$0.63	&$+$0.77	&$-$0.21	&$+$0.98	&$-$0.14 &\\
J22492756$-$2238289	&$-$1.82	&$-$0.08	&$+$0.39	&$-$0.43	&$+$0.48	&$-$0.91	&$+$0.82 &$r$-I\\
J22531984$-$2248555	&$-$1.87	&$+$0.89	&$+$0.55	&$+$1.01	&$+$0.88	&$+$0.13	&$-$0.46 &\\
J22595884$-$1554182	&$-$2.64	&$-$0.86	&$+$0.03	&$-$0.58	&$<-$0.30	&$>-$0.28	&$+$0.61 &limited-$r$\\
J23022289$-$6833233	&$-$2.64	&$+$0.30	&$+$0.15	&$-$0.29	&$+$0.61	&$-$0.90	&$+$0.44 &$r$-I\\
J23044914$-$4243477	&$-$2.43	&$-$0.62	&$+$0.48	&$-$0.42	&$<+$0.07	&$>-$0.49 	&$+$0.90 &limited-$r$\\
J23130003$-$4507066	&$-$2.64	&$+$0.28	&$-$0.60	&$-$0.65	&$<+$0.05	&$>-$0.70	&$+$0.05 &Unknown\\
J23265258$-$0159248	&$-$3.13	&$-$0.63	&$+$0.07	&$+$0.31	&$+$0.69	&$-$0.38	&$-$0.24 &$r$-I\\
J23310716$-$0223301	&$-$2.28	&$+$1.95	&$+$0.53	&$+$2.23	&$+$2.09	&$+$0.14	&$-$1.70 &CEMP-$s$\\
J23362202$-$5607498	&$-$2.06	&$-$0.05	&$>+$1.00	&$+$0.24	&$+$1.14	&$-$0.90	&$>+$0.76 &$r$-II\\
\enddata		     					      
\end{deluxetable*}  

Final abundances derived for the 107 stars are listed in Table \ref{tab:abun},
along with their classifications based on these abundances. Table
\ref{tab:error} lists the abundance uncertainties arising from 
stellar-parameter uncertainties, along with random errors, for three stars representative of the stellar-parameter range of our sample. The random error is dominated by the uncertainty in continuum placement, and is estimated by deriving abundances for the lines with $\pm1\sigma$ placement of continuum, based on our average S/N$=30$. We computed these uncertainties by deriving abundances using atmosphere models varied by the uncertainty of the different parameters as quoted in \S\ref{analysis}. These uncertainties are then added in quadrature, along with an additional estimated 0.05~dex random error.

\begin{deluxetable*}{lcccccc}
\tablecaption{Uncertainties for Three Representative Stars in the Sample \label{tab:error}}
\tablewidth{0pt}
\tablecolumns{7}
\tablehead{ \colhead{Abundance} & \colhead{$T_{\rm eff}$} &
  \colhead{$\log g$} &
  \colhead{$\sigma_\mathrm{[Fe/H]}$} & \colhead{$\xi$} &
  \colhead{$\sigma_{\rm rand}$} & \colhead{$\sigma_{\rm tot}$}\\
\colhead{} & \colhead{($\pm$150~K)} & \colhead{($\pm$0.3~dex)} &\colhead{(dex)} &\colhead{($\pm$0.3~dex)} &\colhead{(dex)} &\colhead{(dex)} }
\startdata 
\cutinhead{J02500719$-$5145148 (4707/1.40/$-$2.20/2.30)}
$\mathrm{[C/Fe]}$ & $\pm$0.30 & $\pm$0.04 & $\pm$0.21&$\pm$0.01 & $\pm$0.05 &$\pm$0.37\\
$\mathrm{[Sr/Fe]}$& $\pm$0.15 & $\pm$0.08 & $\pm$0.21&$\pm$0.22 & $\pm$0.05 &$\pm$0.35 \\
$\mathrm{[Ba/Fe]}$& $\pm$0.12 & $\pm$0.13 & $\pm$0.21&$\pm$0.24 & $\pm$0.05 &$\pm$0.37 \\
$\mathrm{[Eu/Fe]}$& $\pm$0.13 & $\pm$0.08 & $\pm$0.21&$\pm$0.02 & $\pm$0.05 &$\pm$0.26 \\
\cutinhead{J13164824$-$2743351 (4990/2.24/$-$1.61/1.50)}
$\mathrm{[C/Fe]}$ & $\pm$0.28 & $\pm$0.09 & $\pm$0.18&$\pm$0.04 & $\pm$0.05 &$\pm$0.35\\
$\mathrm{[Sr/Fe]}$& $\pm$0.08 & $\pm$0.08 & $\pm$0.18&$\pm$0.05 & $\pm$0.05 &$\pm$0.22 \\
$\mathrm{[Ba/Fe]}$& $\pm$0.08 & $\pm$0.14 & $\pm$0.18&$\pm$0.24 & $\pm$0.05 &$\pm$0.34 \\
$\mathrm{[Eu/Fe]}$& $\pm$0.07 & $\pm$0.12 & $\pm$0.18&$\pm$0.07 & $\pm$0.05 &$\pm$0.24 \\
\cutinhead{J20514971$-$6158008 (5285/3.13/$-$1.87/1.68)}
$\mathrm{[C/Fe]}$  & $\pm$0.28 & $\pm$0.03 & $\pm$0.15&$\pm$0.01 & $\pm$0.05 &$\pm$0.32\\
$\mathrm{[Sr/Fe]}$ & $\pm$0.13 & $\pm$0.09 & $\pm$0.15&$\pm$0.07 & $\pm$0.05 &$\pm$0.23 \\
$\mathrm{[Ba/Fe]}$ & $\pm$0.09 & $\pm$0.09 & $\pm$0.15&$\pm$0.09 & $\pm$0.05 &$\pm$0.22 \\
$\mathrm{[Eu/Fe]}$ & $\pm$0.08 & $\pm$0.11 & $\pm$0.15&$\pm$0.03 & $\pm$0.05 &$\pm$0.21 \\
\enddata		     					      
\end{deluxetable*} 

\subsection{Radial-Velocity Variations}

For nine of the stars in our sample, we find velocities differing from those reported by the RAVE survey DR5 \citep{kunder2017}; these are listed in Table \ref{tab:binary}. The typical uncertainty of the radial velocities listed by RAVE DR5 is $< 2$ km~s$^{-1}$ \citep{kunder2017}; the uncertainties of our radial velocities are typically $< 1$ km~s$^{-1}$ (see Table \ref{tab:obslog}). We have therefore chosen to highlight stars for which the two measurements differ by more than 5 km~s$^{-1}$. One of these, J225319.9$-$224856, is a CEMP-$s$ star, a class of stars enhanced in both C and $s$-process elements, and primarily found in binary systems \citep{hansen2016}. Four of these stars are $r$-II stars, while two are $r$-I stars, resulting in
six possible binaries among the 50 new $r$-I and $r$-II stars discovered in our sample. Additionally, one of the $r$-II star rediscoveries, HE~1523$-$0901, is also known to be in a binary system, while the other, CS~31082$-$001, shows no sign of radial velocity variation \citep{hansen2015b}. No binary information is available for the $r$-I rediscoveries. In total, this results in a binary frequency of 11.1$\pm$4.3\% (assuming binomial errors), which agrees with the binary frequency of 18$\pm$6\% previously found for $r$-process-enhanced stars
\citep{hansen2011,hansen2015b}. This is a lower limit, given that, e.g.,
HE~1523-0901 has radial-velocity variations well below 1\,km~s$^{-1}$, a level that is not accounted for here. However, it is striking that four of our ten newly-detected $r$-II stars are apparent binaries, whereas \citet{hansen2015b} only found one binary system among the nine $r$-II stars included in their sample. We note that none of the possible binary systems identified in Table \ref{tab:binary} are double-lined spectroscopic binaries.

\begin{deluxetable}{lrrrl}
\tablecaption{Radial Velocities from RAVE DR5 and our Data for Possible Binaries. \label{tab:binary}}
\tabletypesize{\scriptsize}
\tablewidth{0pt}
\tablehead{
\colhead{Stellar ID} & \colhead{MJD} & \colhead{RV} & \colhead{RV$_{\rm err}$} & \colhead{Ref.}\\
\colhead{} & \colhead{} & \colhead{(km~s$^{-1}$)} & \colhead{(km~s$^{-1}$)} & \colhead{}}
\startdata 
J02040793$-$3127556 &  55866.62968  & $+$89.0  &     0.5 & RAVE DR5\\
   					&  57607.90877  & $+$83.4  &     1.0 & this work \\
J02355867$-$6745520 &  54424.55569  &$+$112.5  &     2.1 & RAVE DR5 \\
   ($r$-I)          &  57610.82171  &$+$101.5  &     0.2 & this work\\
J02462013$-$1518419 &  55804.65655  &$+$287.5  &     2.5 & RAVE DR5\\
	($r$-II)		&  57609.92227  &$+$278.2  &     0.4 & this work  \\
J17225742$-$7123000 &  55706.67402  &$+$284.3  &     2.5 & RAVE DR5 \\
	($r$-II) 		&  57605.55568  &$+$277.3  &     0.5 & this work \\
			  		&  57607.54850  &$+$278.1  &	 1.1 & this work  \\
J17400682$-$6102129 &  55441.36433  &$+$262.4  &     0.5 & RAVE DR5 \\
	($r$-I)	   		&  57606.57576  &$+$255.0  &     0.3 & this work   \\
J19534978$-$5940001 &  55687.77159  &$+$209.3  &     1.5 & RAVE DR5\\
	(limited-$r$)		&  57608.66299  &$+$216.4  &     0.6 & this work \\
J20093393$-$3410273 &  54952.76815  & $+$33.6  &     1.0 & RAVE DR5\\
	($r$-II)   		&  57607.68078  & $+$27.2  &     0.3 & this work \\
J21064294$-$6828266 &  55716.75769  & $-$64.9  &     2.6 & RAVE DR5\\
	($r$-II)		&  55738.76316  & $-$71.6  &    18.7 & RAVE DR5\\
			  		&  57608.73613  & $-$72.7  &     0.6 & this work \\
J22531984$-$2248555 &  53923.72365  & $+$12.3  &     0.9 & RAVE DR5\\
	(CEMP-$s$) 		&  57607.78387  &  $-$7.0  &     0.1 & this work\\
\enddata 
\end{deluxetable}

\section{Discussion}

\subsection{Newly Identified $r$-Process-Enhanced Stars}

Following the abundance-selection criteria listed in Table
\ref{tab:selection}, we identify 12 $r$-II, 39 $r$-I, 20 limited-$r$, and 3 CEMP-$r$ stars among the 107 stars observed. The three CEMP-$r$ stars all have an Eu enhancement in the $r$-I star Eu range, we therefore include them in this group in the following figures and discussion. Two of the identified $r$-II stars are rediscoveries: J15260106$-$0911388 = HE~1523$-$0901 \citep{frebel2007}, and J01293113$-$1600454 = CS~31082$-$001 \citep{hill2002}. Two of the $r$-I stars are rediscoveries: J21063474$-$4957500
= HD~200654 \citep{roederer2014a} and J23265258$-$0159248
\citep{thanathibodee2016}. Thus, we have identified ten new $r$-II stars and 37 new $r$-I stars. The frequencies of new $r$-process-enhanced stars are somewhat higher than was found by \citet{barklem2005}, who identified 8 $r$-II and 35 $r$-I stars in their sample of 253 stars. However, the stars in the sample of \citet{barklem2005} are generally warmer and fainter than our targets, often impeding the detection of Eu and other neutron-capture elements if the target S/N is not reached for the data. Hence, a number of their non-detections may not be real, but rather, a result of higher stellar temperature, low S/N, or a combination of both. 

Specific searches for stars that exhibit a limited $r$-process signature have not been carried out in the past, although multiple works have recognized the existence of these stars \citep[e.g.,][]{travaglio2004,honda2006,jacobson2015}. We find 20 stars with abundances satisfying our limited-$r$ criteria (see Table \ref{tab:selection}), which is about half of the stars with low Eu abundances in our sample. 

We note that four CEMP-$s$ stars were also found in our sample, again because not all targets for this run had been through our eventual Phase-I vetting process. 

\subsection{Metallicity Distribution}

\begin{figure}
\center
\includegraphics[scale=0.45]{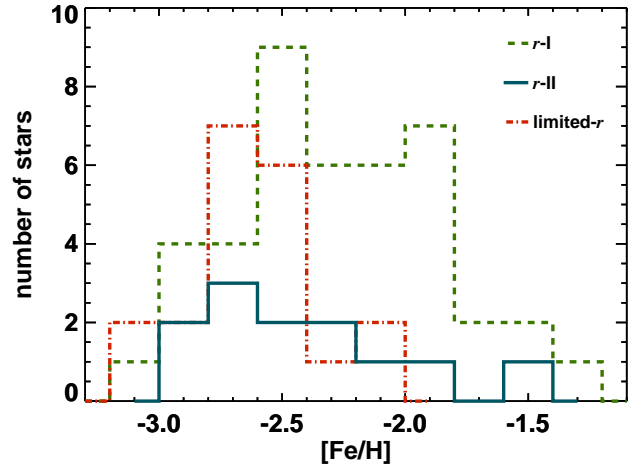}
\caption{Histogram showing the number of detected $r$-I (dashed green line), $r$-II (solid blue line), and limited-$r$ (dot-dashed red line) stars in our sample (including duplicates), as a function of metallicity. \label{fig:hist}}  
\end{figure}

Figure \ref{fig:hist} shows the number of identified $r$-I, $r$-II, and
limited-$r$ stars as a function of metallicity. We detect
both $r$-I and $r$-II stars over a wide metallicity range ($-3.0 \lesssim \mathrm{[Fe/H]} \lesssim -1.5$). The limited-$r$ stars are found primarily at the low-metallicity end; from $\mathrm{[Fe/H]}\sim -2$ and down to $\mathrm{[Fe/H]} \sim -3$. The observed distribution for the limited-$r$ stars indicates that this signature is more clear in metal-poor environments, where chemical evolution has not yet erased signatures of individual processes, as opposed to more chemically-evolved systems, in which only processes leaving a strong chemical signature can be identified anymore, like the $r$-II stars. However, the currently available sample remains small. With future, larger samples of $r$-I, $r$-II, and limited-$r$ stars collected by the RPA, we will soon be able to better determine the metallicity distributions of these stars.

The metallicity distribution of the our sample also differs from that of \citet{barklem2005}. The \citet{barklem2005} stars have a median $\mathrm{[Fe/H]} \sim -2.8$, after selecting candidates with $\mathrm{[Fe/H]}
\le -2.5$. In contrast, we chose $\mathrm{[Fe/H]} < -2.0$ for our initial candidate selection, and the median of our sample is $\mathrm{[Fe/H]}\sim -2.4$. All of the $r$-II stars detected by \citet{barklem2005} were found in a narrow metallicity range around $\mathrm{[Fe/H]}\sim -3$, as do many of the $r$-II stars discovered since. However, we detect $r$-II stars in a wider metallicity range, from $\mathrm{[Fe/H]}= -3$ to $\mathrm{[Fe/H]}= -1.5$, suggesting that the previous metallicity preference of these stars is likely an artifact or selection biases toward the overall lower metallicity in previous samples. 

Interestingly, the metallicity distribution of the $r$-I and $r$-II stars found in the Milky Way satellites is more similar to what we find \citep{shetrone2001,letarte2010}. The one $r$-II star found in the bulge also has $\mathrm{[Fe/H]}\sim -2$ \citep{johnson2013}. Furthermore, we recently detected an $r$-II star with $\mathrm{[Fe/H]} = -2$ in our Northern Hemisphere sample \citep{sakari2018}.

Our sample also includes the most metal-rich $r$-II star detected in the Milky Way halo to date, J18024226$-$4404426, with $\mathrm{[Fe/H]} = -1.55$, $\mathrm{[Eu/Fe]} = +1.05$ and $\mathrm{[Ba/Eu]} = -0.10$. The relatively high $\mathrm{[Ba/Eu]}$ ratio found in J18024226$-$4404426 clearly shows that a substantial contribution from the $s$-process to the neutron-capture element abundances is present at this metallicity. Nevertheless, the neutron-capture-element abundance pattern of this star is dominated by the $r$-process, and, together with the other $r$-II stars discovered in this search, it widens the otherwise narrow metallicity distribution of the known $r$-II stars in the halo.  

\subsection{Magnitude Distribution}
\begin{figure}
\center
\includegraphics[scale=0.45]{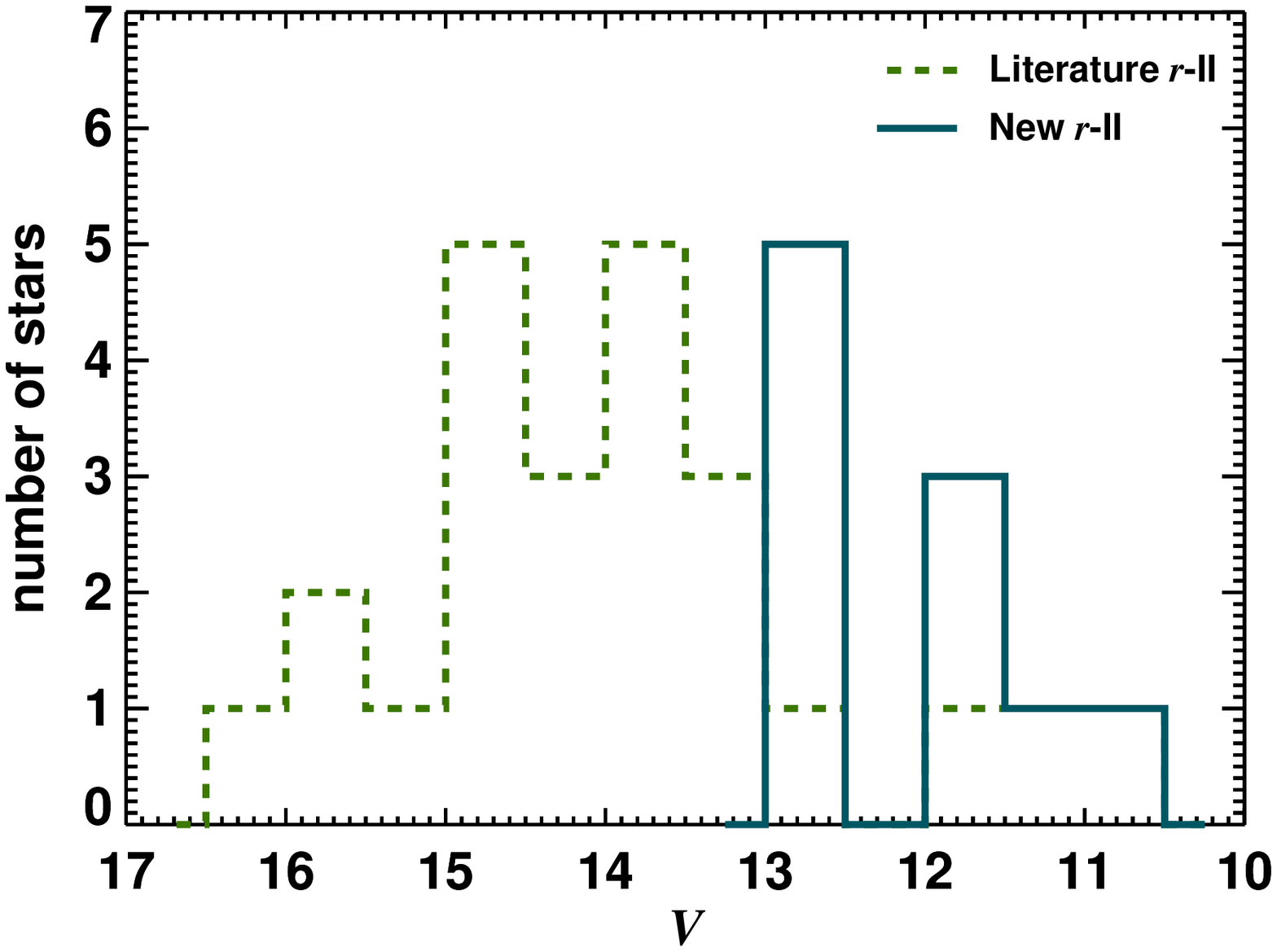}
\caption{$V$ magnitudes for the literature sample of halo $r$-II stars and the new $r$-II stars found in our sample \citep{rossi2005,beers2007,zacharias2012,cohen2013,munari2014,roederer2014a} \label{fig:histV}}  
\end{figure}

To demonstrate the difference in $V$ magnitudes for our ten newly-discovered $r$-II stars compared to the literature sample $r$-II stars in the halo, we have plotted the number of $r$-II stars as a function of $V$ magnitude in Figure \ref{fig:histV} for our stars and the literature stars for which we could find $V$ magnitudes. All of our new $r$-II
stars have $V < 13$. Only four $r$-II stars this bright were known previously (J203843.2-002333, $V=12.7$, \citealt{placco2017}; CS~31082$-$001, $V=11.6$, \citealt{hill2002}; HE~1523$-$0901, $V=11.1$, \citealt{frebel2006}; and J153830.9$-$180424, $V = 10.86$, \citealt{sakari2018}). 

This sample also includes the brightest $r$-II star detected in the Milky Way's halo to date, J21091825$-$1310062, with $V = 10.7$, $\mathrm{[Fe/H]} = -2.40$, $\mathrm{[Eu/Fe]} = +1.25$, and $\mathrm{[Ba/Eu]} = -1.13$. This star has one of the lowest
$\mathrm{[Ba/Eu]}$ ratios found in our sample, indicating a pure $r$-process origin for the neutron-capture elements seen in this star. Note that the second brightest $r$-II star was discovered in the northern search for $r$-process-enhanced stars conducted by the RPA \citep{sakari2018}. As the present manuscript was being prepared, yet another $r$-II star, even brighter than either of the above two stars ($V = 10.1$), was identified from Northern Hemisphere snapshot spectroscopy with the McDonald 2.7m telescope.  A portrait spectrum was also obtained, and is reported on by Holmbeck et al. (in prep.).  The brightness of the $r$-II stars in our sample means that we will be able to derive very complete neutron-capture-element abundance patterns for them, including thorium and possibly uranium, which is essential to distinguish between different $r$-process models.  A number are sufficiently bright to be studied in the near-UV at high spectral resolution with the Hubble Space Telescope. 

\subsection{Sr, Ba, Eu as $R$-Process Indicators}
\begin{figure}
\center
\includegraphics[scale=0.45]{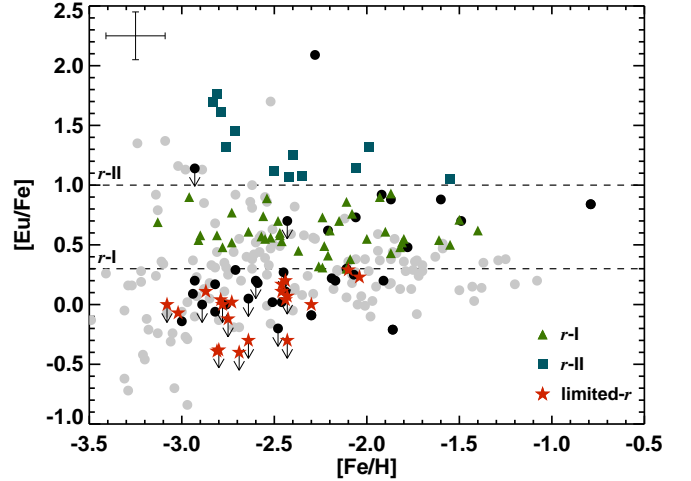}
\caption{Derived $\mathrm{[Eu/Fe]}$ abundances for the sample, as a function of metallicity:  $r$-I stars (green triangles), $r$-II stars (blue squares), limited-$r$ (red stars), and non $r$-process-enhanced stars (black dots); upper limits are shown with black arrows. For comparison, Eu measurements from \citet{roederer2014a} are shown as gray dots. Division lines are drawn for non-$r$-process-enhanced, $r$-I, and $r$-II stars, shown with dashed lines at $\mathrm{[Eu/Fe]} = +0.3$ and $+1.0$. \label{fig:EuFe}} 
\end{figure}

With the stellar-parameter range and S/N of this sample, we are able to detect Eu abundances in our stars down to $\mathrm{[Eu/Fe]} \sim -0.2$. The distribution of Eu abundances for our sample stars as a function of
metallicity is shown in Figure \ref{fig:EuFe}. For comparison, we have also included Eu measurements from the large sample of
\citet{roederer2014a}. Similar to \citet{roederer2014a}, we find a large
spread ($>2$~dex) in the Eu abundances derived for our sample. This spread has been interpreted as a possible sign of multiple production sites for $r$-process elements \citep{sneden2000,travaglio2004}. In particular, as we are able to detect Eu in stars over the full metallicity range of our sample, indicating the need for a source of $r$-process elements early in the Universe.  

From Figure \ref{fig:EuFe}, it can also be seen that two of our identified limited-$r$ stars have sub-Solar $\mathrm{[Eu/Fe]}$ abundances, a number of them cluster around around the Solar value, and some are mildly enhanced in Eu. Our template star for the limited-$r$ stars, HD~122563, has sub-Solar $\mathrm{[Sr/Fe]}$, $\mathrm{[Ba/Fe]}$, and $\mathrm{[Eu/Fe]}$, and is thus $r$-element poor. We have chosen to include more stars in this group, requiring that they show lower Eu abundances than the $r$-I (and $r$-II) stars and larger abundances of the light $r$-process elements, compared to the heavy $r$-process elements ($\mathrm{[Sr/Ba]} > +0.5$), as we wish to
investigate the production of the light $r$-process elements with this group of stars.

\begin{figure}
\center
\includegraphics[scale=0.45]{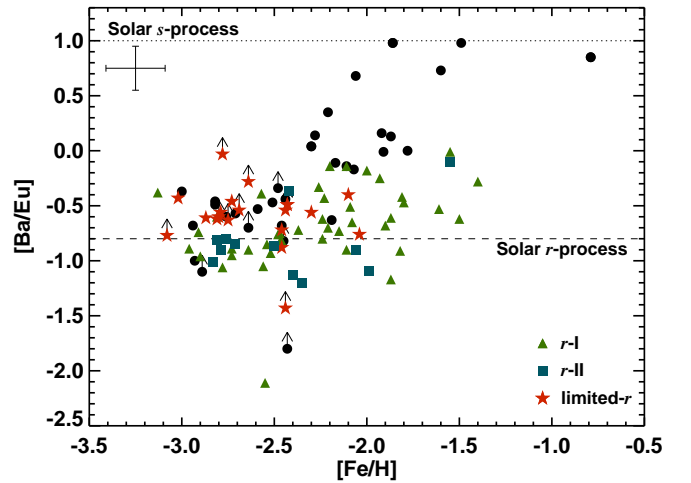}
\caption{Derived $\mathrm{[Ba/Eu]}$ abundances for the sample, as a function of metallicity: $r$-I stars (green triangles), $r$-II stars (blue squares), limited-$r$ (red stars), and non-$r$-process-enhanced stars (black dots); lower limits are shown with gray arrows. The dotted line is the Solar System $s$-process fraction ($\mathrm{[Ba/Eu]} = +1.0$), the dashed line is the Solar System $r$-process fraction ($\mathrm{[Ba/Eu]} = -0.8$), both taken from \citet{simmerer2004} \label{fig:EuBaFe}}  
\end{figure}

The $\mathrm{[Ba/Eu]}$ ratio is often used as a diagnostic for the level of $r$- versus $s$-process origin of the neutron-capture elements measured in stars. Figure \ref{fig:EuBaFe} shows the $\mathrm{[Ba/Eu]}$ ratios for our sample stars, as a function of metallicity, along with the Solar System $r$- and $s$-process ratios from \citet{simmerer2004}. The majority of our newly discovered $r$-II stars have $\mathrm{[Ba/Eu]} < -0.8$, indicating a pure $r$-process origin for the neutron-capture-element content of these stars. The $r$-I and limited-$r$ stars exhibit a larger variation in their $\mathrm{[Ba/Eu]}$ values, slightly increasing with metallicity. 

\begin{figure}
\center
\includegraphics[scale=0.45]{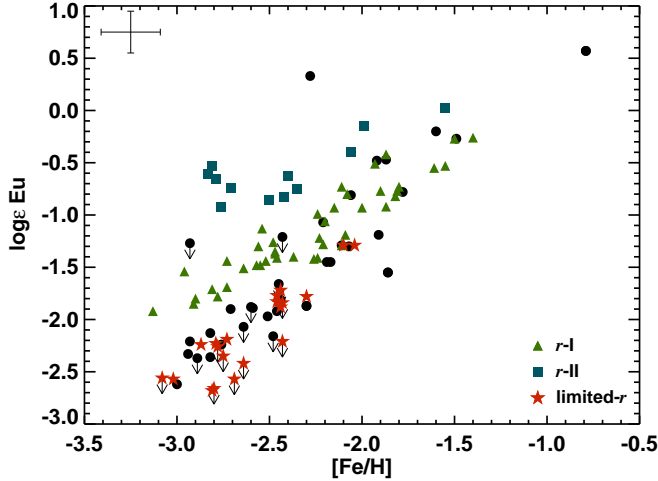}
\caption{Absolute Eu abundances for the sample, as a function of metallicity: $r$-I stars (green triangles), $r$-II stars (blue squares), limited-$r$ (red stars), and non-$r$-process-enhanced stars (black dots); upper limits are shown with gray arrows. \label{fig:logEuFe}}  
\end{figure}

We also plot the absolute Eu abundances derived for our stars, as a function of metallicity, in Figure \ref{fig:logEuFe}. A clear increase of Eu with Fe is seen for both the $r$-process-enhanced and non-$r$-process-enhanced stars; a similar trend is also seen for $r$-process-enhanced stars detected in dwarf galaxies \citep{hansen2017}. The slopes for the $r$-I and $r$-II stars appear to differ in this plot, indicating different enrichment of these stars. It is likely that the sampling of $r$-I stars at the low-metallicity end is incomplete, due to the difficulty in detecting the lines for low Eu abundances in these stars. Also, the sampling of $r$-II stars at higher metallicity is sparse. Hence, the difference in slopes may just be a result of the metallicity sampling difference. This will be easier to determine with future larger samples of $r$-I and $r$-II stars gathered by the RPA.

\begin{figure}
\center
\includegraphics[scale=0.45]{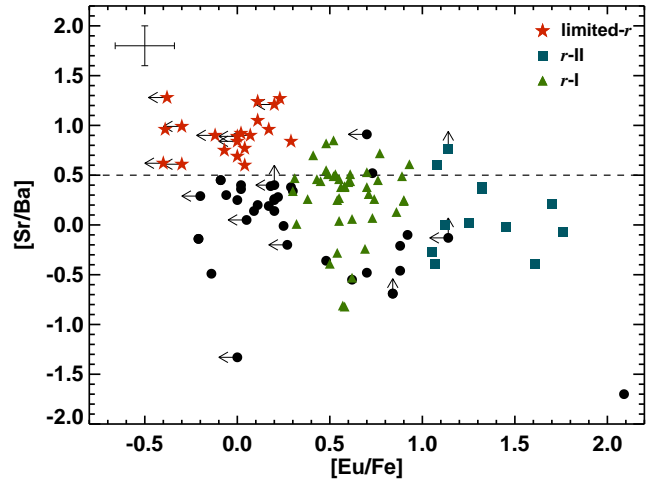}
\caption{Derived $\mathrm{[Sr/Ba]}$ abundances for the sample, as a function of $\mathrm{[Eu/Fe]}$ abundances: $r$-I stars (green triangles), $r$-II stars (blue squares), limited-$r$ (red stars), and non-$r$-process-enhanced stars (black dots); limits are shown with black arrows. Dashed line marks $\mathrm{[Sr/Ba]} = +0.5$.} \label{fig:SrBa}
\end{figure}

Finally, we plot the $\mathrm{[Sr/Ba]}$ abundances, as a function of Eu
abundance, in Figure \ref{fig:SrBa}. The limited-$r$ stars cluster in the upper-left corner of this plot, showing higher abundances of the light $r$-process element Sr than the heavy $r$-process elements Ba and Eu. A few of the $r$-II and $r$-I also exhibit $\mathrm{[Sr/Ba]} > +0.5$, but generally these stars have lower $\mathrm{[Sr/Ba]}$ ratios than the limited-$r$ stars. 

\section{Conclusions}
\label{conclusion}
This paper presents the first data sample from the Southern Hemisphere search for $r$-process-enhanced metal-poor stars in the Galactic halo, conducted as part of the $R$-Process Alliance in an effort to better understand the nature of the $r$-process. We have observed and analyzed a sample of 107 stars, identifying ten new $r$-II stars (12 in total), 40 new $r$-I stars (42 in total, including the CEMP-$r$ stars), and 20 new limited-$r$ stars. Our sample has a wide metallicity range, from $\mathrm{[Fe/H]}= -3.13$ to $-$0.79. While the majority of our newly discovered $r$-II stars are found at the metal-poor end of our sample, we identify these stars all the way up to $\mathrm{[Fe/H]} = -1.5$. Specifically, in this sample we have detected the most metal-rich $r$-II star and the brightest $r$-II star known in the halo today. This metallicity distribution also agrees with that found for stars in dwarf galaxies, especially in the dwarf spheroidal galaxies. The $r$-I stars are detected at all metallicities as well, similar to what has been previously reported for these stars \citep{barklem2005}. In this sample, we find limited-$r$ stars only below $\mathrm{[Fe/H]}\sim -2$. The low $\mathrm{[Ba/Eu]}$ ratios derived for the majority of our detected $r$-II stars suggests a pure $r$-process origin of the neutron-capture elements observed in these stars. The $r$-I and limited-$r$ stars exhibit higher $\mathrm{[Ba/Eu]}$, suggesting a more diluted and/or otherwise mixed
neutron-capture-element content of their natal gas clouds, but still dominated by $r$-process material. Portrait spectra of the majority of the new $r$-II stars reported in this paper have been obtained, and analyses of these, which will enable us to identify possible actinide-boost stars, will be published in upcoming papers.

\acknowledgments{We thank the referee for helpful comments. This publication is based upon work supported in part by the National Science Foundation under grants AST-1108811 and AST-1714873. E.M.H, T.C.B, V.M.P, I.U.R, and A.F. acknowledge partial support from grant PHY~14-30152 (Physics Frontier Center/JINA-CEE), awarded by the U.S.\ National Science Foundation (NSF). A.F. are supported by NSF CAREER grant AST-1255160.
C.M.S. acknowledges funding from the Kenilworth Fund of the New York Community Trust.  This research has made use of NASA's Astrophysics Data System Bibliographic Services. }

\vspace{5mm}
\facilities{du Pont}
\software{MOOG \citep{sneden1973,sobeck2011}, IRAF \citep{tody1986,tody1993} , linemake (https://github.com/vmplacco/linemake), ATLAS9 \citep{castelli2003} }

\bibliographystyle{aa}
\bibliography{THansen}

\end{document}